\documentclass[review,12pt]{elsarticle}

\usepackage{setspace}
\usepackage{subcaption}
\usepackage{multirow}
\usepackage{multicol}
\usepackage{graphicx}
\RequirePackage{amsmath} 
\usepackage{bm}
\usepackage{xfrac}
\usepackage{float}
\usepackage{arydshln} 

\usepackage{amsmath,esint,amsfonts}

\usepackage[most]{tcolorbox}

\usepackage{lineno,hyperref}
\usepackage{cleveref} 
\modulolinenumbers[5]

\journal{Journal of Manufacturing Processes}

\biboptions{sort&compress} 

\begin{document}

\begin{frontmatter}

\title{Optimization of Solidification in Die Casting using Numerical Simulations and Machine Learning}

\author{Shantanu Shahane\fnref{Corresponding Author}}
\author{Narayana Aluru} \author{Placid Ferreira} \author{Shiv G Kapoor} \author{Surya Pratap Vanka}
\address{Department of Mechanical Science and Engineering\\
	University of Illinois at Urbana-Champaign \\
	Urbana, Illinois 61801}
\fntext[Corresponding Author]{Corresponding Author Email: \url{shahaneshantanu@gmail.com}}


%

\begin{abstract}
In this paper, we demonstrate the combination of machine learning and three dimensional numerical simulations for multi--objective optimization of low pressure die casting. The cooling of molten metal inside the mold is achieved typically by passing water through the cooling lines in the die. Depending on the cooling line location, coolant flow rate and die geometry, nonuniform temperatures are imposed on the molten metal at the mold wall. This boundary condition along with the initial molten metal temperature affect the product quality quantified in terms of micro-structure parameters and yield strength. A finite volume based numerical solver is used to determine the temperature-time history and correlate the inputs to outputs. The objective of this research is to develop and demonstrate a procedure to obtain the initial and wall temperatures so as to optimize the product quality. The non-dominated sorting genetic algorithm (NSGA--II) is used for multi--objective optimization in this work. The number of function evaluations required for NSGA--II can be of the order of millions and hence, the finite volume solver cannot be used directly for optimization. Therefore, a multilayer perceptron feed--forward neural network is first trained using the results from the numerical solution of the fluid flow and energy equations and is subsequently used as a surrogate model. As an assessment, simplified versions of the actual problem are designed to first verify results of the genetic algorithm. An innovative local sensitivity based approach is then used to rank the final Pareto optimal solutions and select a single best design.
\end{abstract}

\begin{keyword}
Die Casting, Deep Neural Networks, Multi--Objective Optimization
\end{keyword}

\end{frontmatter}
\section{Introduction}
Die casting is one of the popular manufacturing processes in which liquid metal is injected into a permanent metal mold and solidified. Generally, die casting is used for parts made of aluminum and magnesium alloys with steel molds. Automotive and housing industrial sectors are common consumers of die casting. In such a complex process, there are several input parameters which affect the final product quality and process efficiency. With advances in computing hardware and software, the physics of these processes can be modeled using numerical simulation techniques. Detailed flow and temperature histories, micro-structure parameters, mechanical strength etc. can be estimated from these simulations. In today's competitive industrial world, estimating the values of input parameters for which the product quality is optimized has become highly important. There has been extensive research in numerical optimization algorithms which can be coupled with detailed numerical simulations in order to handle complex optimization problems.
\par Solidification in casting process has been studied by many researchers. \citet{minaie1991analysis} have analyzed metal flow during die filling and solidification in a two dimensional rectangular cavity. The flow pattern during the filling stage is predicted using the volume of fluid (VOF) method and enthalpy equation is used to model the phase change with convection and diffusion inside the cavity. They have studied the effect of gate location on the residual flow field after filling and the solid liquid interface during solidification. \citet{im2001unified} have done a combined filling and solidification analysis in a square cavity using the implicit filling algorithm with the modified VOF together with the enthaply formulation. They studied the effect of assisting flow and opposite flow due to different gate positions on the residual flow. They found that the liquid metal solidifies faster in the opposite flow than in the assisting flow situation. \citet{cleary2010short} used the Smoothed Particle Hydrodynamics (SPH) to simulate flow and solidification in three dimensional practical geometries. They demonstrated the approach of short shots to fill and solidify the cavity partially with insufficient metal so that validation can be performed on partial castings. \citet{plotkowski2015estimation} simulated the phase change with fluid flow in a rectangular cavity both analytically and numerically. They simplified the governing equations of the mixture model by scaling analysis followed by an analytical solution and then compared with a complete finite volume solution.
\par Recently, there has been a growing interest in the numerical optimization of various engineering systems. \citet{poloni2000hybridization} applied neural network with multi--objective genetic algorithm and gradient based optimizer to the design of a sailing yacht fin. The geometry of the fin was parameterized using Bezier polynomials. The lift and drag on the fin was optimized as a function of the Bezier parameters and thus, an optimal fin geometry was designed. \citet{elsayed2013cfd} performed a multi--objective optimization of a gas cyclone which is a device used as a gas-solid separator. They trained a radial basis function neural network (RBFNN) to correlate the geometric parameters like diameters and heights of the cyclone funnel to the performance efficiency and the pressure drop using the data from numerical simulations. They further used the non-dominated sorting genetic algorithm (NSGA--II) to obtain the Pareto front of the cyclone designs. \citet{wang2018optimization} optimized the groove profile to improve hydrodynamic lubrication performance in order to reduce the coefficient of friction and temperature rise of the specimen. They coupled the genetic algorithm (GA) with the sequential quadratic programming (SQP) algorithm such that the GA solutions were provided as initial points to the SQP. \citet{stavrakakis2011selection} solved for window sizes for optimal thermal comfort and indoor air quality in naturally ventilated buildings. A computational fluid dynamics model was used to simulate the air flow in and around the buildings and generate data for training and testing of a RBFNN which is further used for constrained optimization using the SQP algorithm. \citet{wei2003optimization} modeled the thermal resistance of a micro-channel heat exchanger for electronic cooling using a simplified thermal resistance network model. They used a genetic algorithm to obtain optimal geometry of the heat exchanger so as to minimize the thermal resistance subject to constraints of maximum pressure drop and volumetric flow rate. \citet{husain2010enhanced} optimized the thermal resistance and pumping power of a micro-channel heat sink as a function of geometric parameters of the channel. They used a three dimensional finite volume solver to solve the fluid flow equations and generate training data for surrogate models. They used multiple surrogate models like response surface approximations, Kriging and RBFNN. They provided the solutions obtained from the NSGA--II algorithm to SQP as initial guesses. \citet{lohan2017topology} performed a topology optimization to maximize the heat transfer through a heat sink with dendritic geometry. They used a space colonization algorithm to generate topological patterns with a genetic algorithm for optimization. \citet{amanifard2008modelling} solved an optimization problem to minimize the pressure drop and maximize the Nusselt number with respect to the geometric parameters and Reynolds number for micro-channels. They used a group method of data handling type neural network as a surrogate model with the NSGA--II algorithm for optimization. \citet{esparza2006optimal} optimized the design of a gating system used for gravity filling a casting so as to minimize the gate velocity. They used a commercial program (FLOW3D) to estimate the gate velocity as a function of runner depth and tail slope and the SQP method for optimization. \citet{gc2018systematic} modeled and optimized the wear behavior of squeeze cast products. Three different optimization methods (genetic algorithm, particle swarm optimization and desirability function approach) are combined with neural network as a surrogate model. The training data for neural network is obtained from experiments and nonlinear regression models.
\par In this paper, we consider the heat transfer and solidification processes in die casting of a complex model geometry. The computer program \cite{shahane2019finite} solves the fluid flow and energy equations, coupled with the solid fraction--temperature relation, using a finite volume numerical method. When not significant, natural convection flow is neglected and only the energy equation is solved. The product quality is assessed using grain size and yield strength which are estimated using empirical relations. The solidification time is used to quantify the process efficiency. The molten metal and mold wall temperatures are crucial in determining the quality of die casting. The wall temperature is typically nonuniform due to the complex mold geometries and asymmetric placement of cooling lines. This nonuniformity can be modeled by domain decomposition of the wall and assigning single temperature value to each domain. Neural networks are trained using the data generated from the simulations to correlate the initial and wall temperatures to the output parameters like solidification time, grain size and yield strength. The optimization problem formulated with these three objectives is then solved using genetic algorithm. The procedure illustrated here can be applied to any practical mold geometry with a complex distribution of wall temperatures.
\section{Numerical Model Description}
The numerical model incorporates the effects of solidification and heat transfer in die casting. Since the common die casting geometries have thin cross-sections, the solidification time is of the order of seconds and hence, the effect of natural convection has been found to be negligible. Thus, the momentum equations of the liquid metal are not solved in this work. The energy equation which can be written in terms of temperature has unsteady, diffusion and latent heat terms.
\begin{equation} 
	\rho C_p \frac{\partial T}{\partial t} = \nabla \bullet (k \nabla T) + \rho L_f \frac{\partial f_s}{\partial t} 
	\label{Eq:Energy}
\end{equation}
where, $T$ is temperature, $\rho$ is density, $C_p$ is specific heat, $k$ is thermal conductivity, $L_f$ is latent heat of fusion, $f_s$ is solid fraction and $t$ is time. The Gulliver-Scheil equation (\ref{Eq:Solid_fraction}) \cite{dantzig2001modeling} relates solid fraction to temperature for a binary alloy.
\begin{equation}
	f_s(T)=
	\begin{cases}
	0 & \text{if } T  >T_{liq}\\
	1 & \text{if } T  <T_{sol}\\
	1 - \left( \frac{T-T_f}{T_{liq} - T_f}\right)^{\frac{1}{k_p-1}}  & \text{otherwise}
	\end{cases}
	\label{Eq:Solid_fraction}
\end{equation}
where, $k_p$ is partition coefficient, $T_f$ is freezing temperature, $T_{sol}$ is solidus temperature and $T_{liq}$ is liquidus temperature.
\par Secondary Dendrite Arm Spacing (SDAS) is a microstructure parameter which can be used to estimate the 0.2\% yield strength. The cooling rate at each point in the domain is computed by numerically solving the energy equation and solid fraction temperature relation (\cref{Eq:Energy,Eq:Solid_fraction}). The following empirical relations link the cooling rate to SDAS and yield strength.
\begin{equation} 
	\text{SDAS } = \lambda_2 =A_\lambda\left(\frac{\partial T}{\partial t}\right)^{B_\lambda} \hspace{10pt} [\text{in }\mu \text{m}].
	\label{Eq:SDAS}
\end{equation}
where, $A_\lambda = 44.6$ and $B_\lambda = -0.359$ are based on the model for microstructure in aluminum alloys \cite{backer2007microporosity}.
\begin{equation} 
	\sigma_{0.2} = A_\sigma \lambda_2 ^{-1/2} + B_\sigma
	\label{Eq:YS}
\end{equation}
where, $\sigma_{0.2}$ is in MPa, $\lambda_2$ (SDAS) is in $\mu$m, $A_\sigma=59.0$ and $B_\sigma=120.3$~\cite{okayasu2015precise}.
\par Grain size estimation is based on the work of \citet{greer2000modelling}. The grain growth rate is given by:
\begin{equation} 
	\frac{dr}{dt} = \frac{\lambda_s^2 D_s}{2 r}
	\label{Eq:grain_growth_rate}
\end{equation}
where, $r$ is the grain size, $D_s$ is the solute diffusion coefficient in the liquid and $t$ is the time. The parameter $\lambda_s$ is obtained using invariant size approximation:
\begin{equation} 
	\lambda_s = \frac{-S}{2 \pi^{0.5}} + \left(\frac{S^2}{4 \pi} - S\right)^{0.5}
	\label{Eq:grain_growth_lambda}
\end{equation}
$S$ is given by 
\begin{equation} 
	S = \frac{2(C_s - C_0)}{C_s - C_l}
	\label{Eq:grain_growth_S}
\end{equation}
where, $C_l = C_0 (1-f_s)^{(k_p-1)}$ is solute content in the liquid, $C_s = k_p C_l$ is solute content in the solid at the solid-liquid interface and $C_0$ is the nominal solute concentration. Hence, from the partition coefficient ($k_p$) and estimated solid fraction ($f_s$), \cref{Eq:grain_growth_rate,Eq:grain_growth_S,Eq:grain_growth_lambda} are solved to get the final grain size.
\begin{figure}[H]
	\centering
	\begin{subfigure}[t]{0.49\textwidth}
		\includegraphics[width=\textwidth]{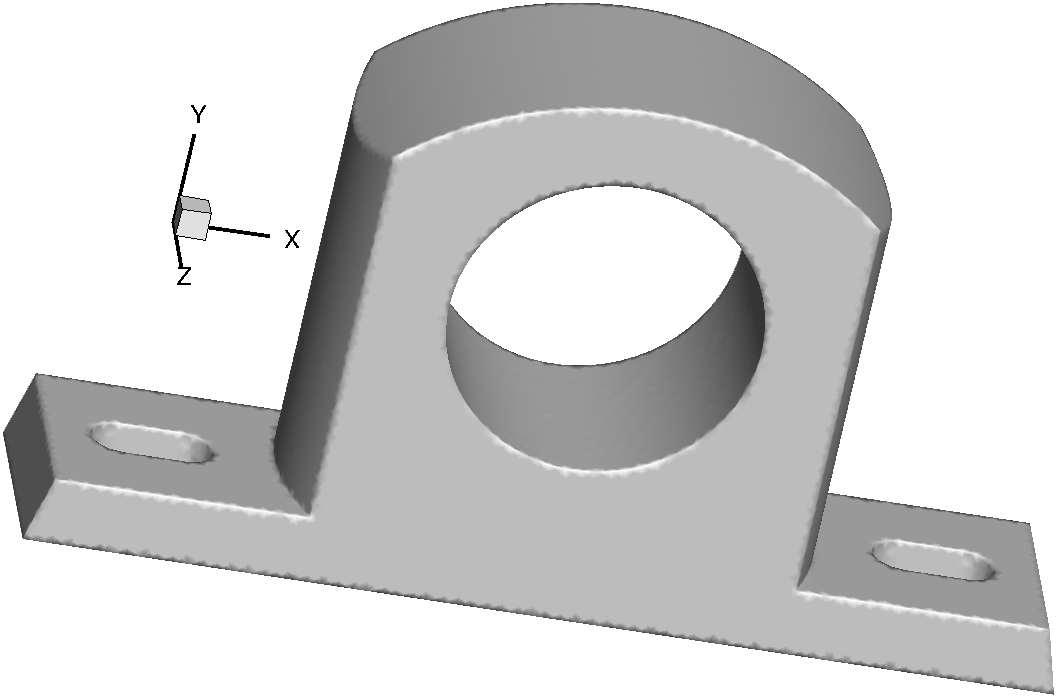}
		\caption{Geometry}
	\end{subfigure}
	\begin{subfigure}[t]{0.49\textwidth}
		\includegraphics[width=\textwidth]{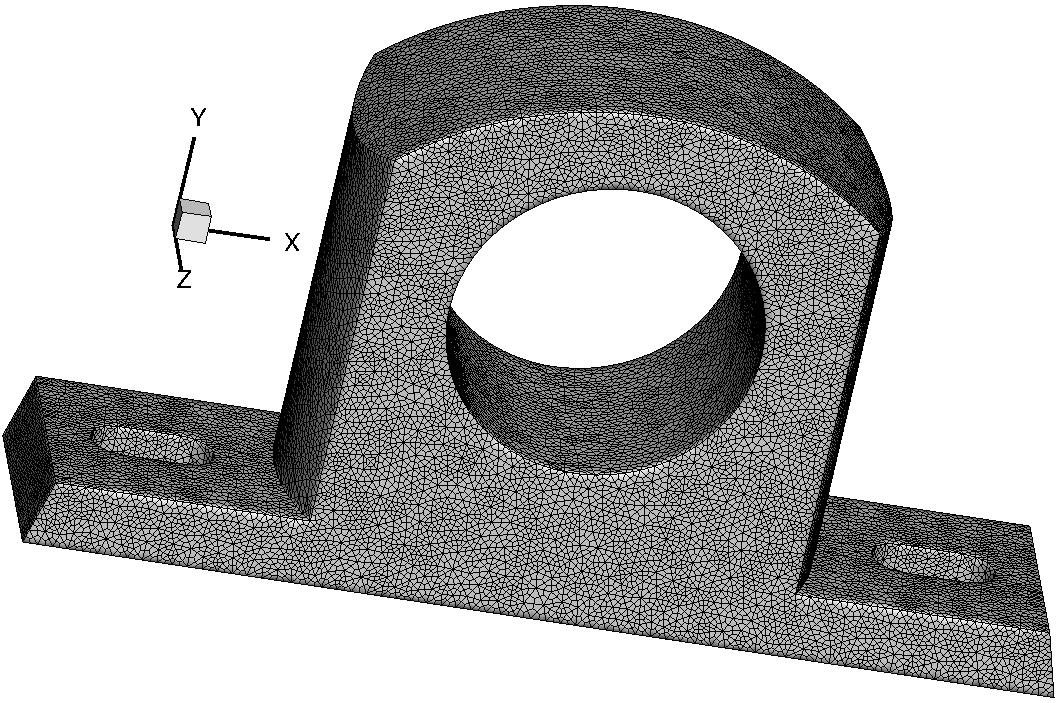}
		\caption{Mesh: 334000 Elements}
	\end{subfigure}	 
	\caption{Clamp: 16.5 cm x 9 cm x 3.7 cm \cite{shahane2019finite}}
	\label{Fig:Clamp}
\end{figure}
\par
\begin{figure}[H]
	\centering
	\begin{subfigure}[t]{0.49\textwidth}
		\includegraphics[width=\textwidth]{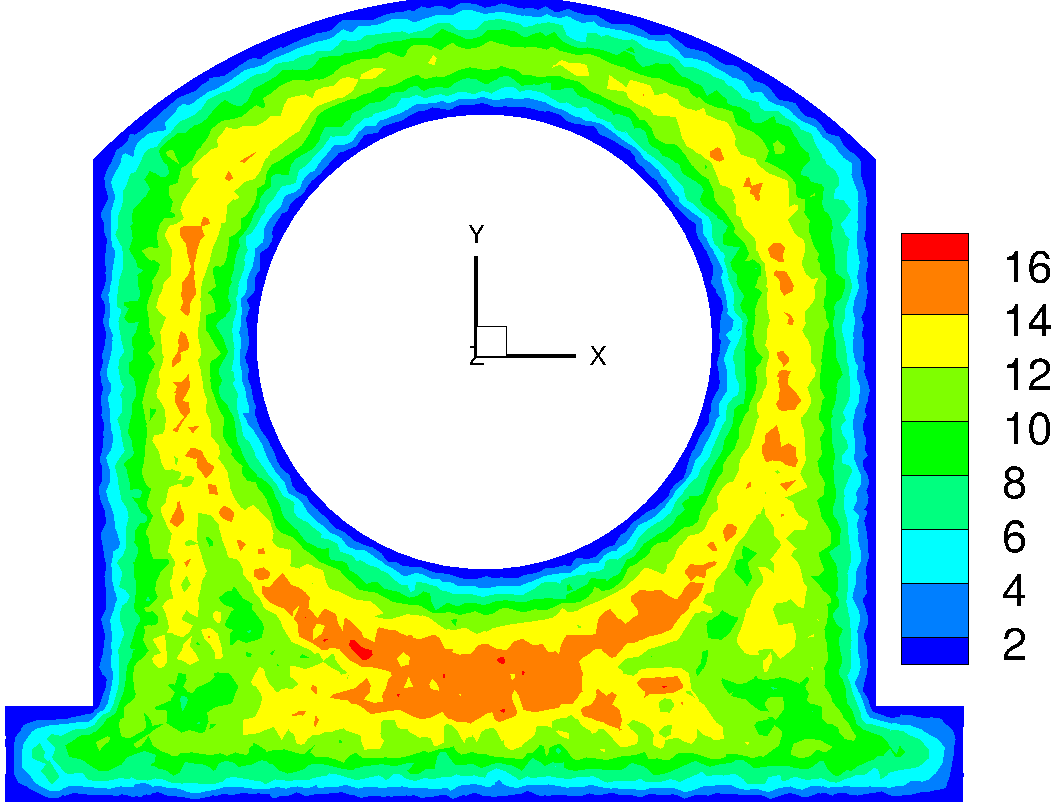}
		\caption{Without Natural Convection
}
	\end{subfigure}
	\begin{subfigure}[t]{0.49\textwidth}
		\includegraphics[width=\textwidth]{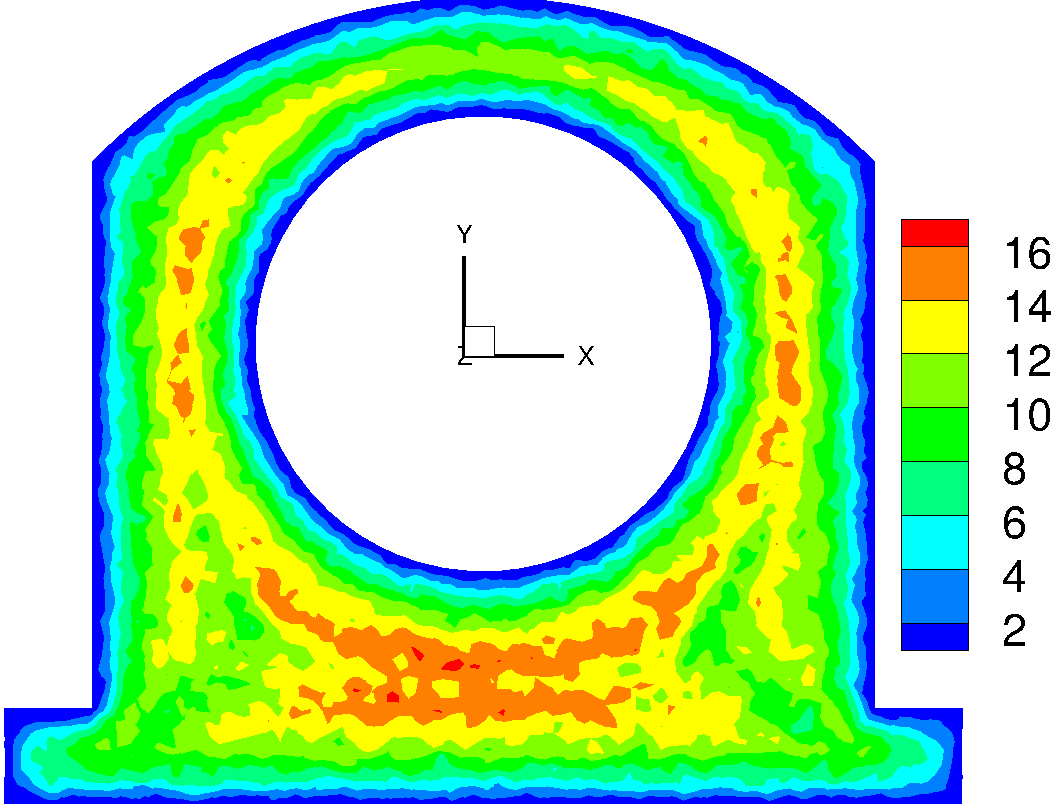}
		\caption{With Natural Convection
}
	\end{subfigure}	 
	\caption{Clamp: Grain Size ($\mu$m)}
	\label{Fig:Clamp Nat Conv Grain Size}
\end{figure}

\begin{figure}[H]
	\centering
	\begin{subfigure}[t]{0.49\textwidth}
		\includegraphics[width=\textwidth]{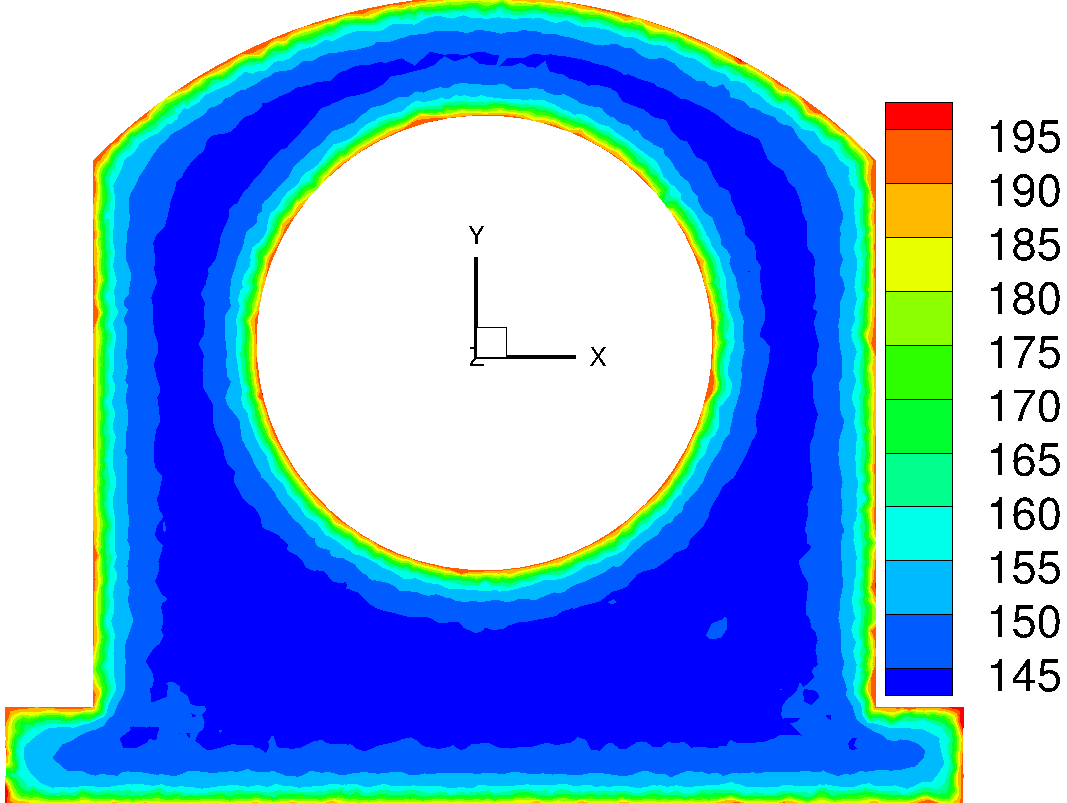}
		\caption{Without Natural Convection
}
	\end{subfigure}
	\begin{subfigure}[t]{0.49\textwidth}
		\includegraphics[width=\textwidth]{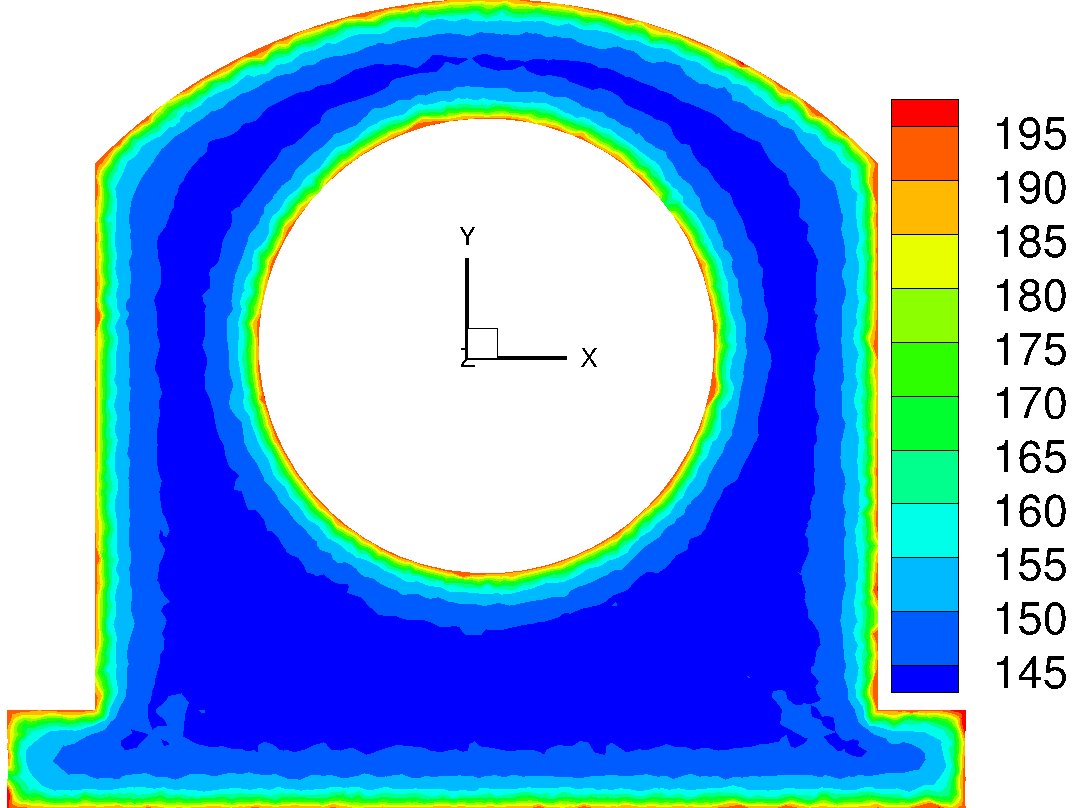}
		\caption{With Natural Convection
}
	\end{subfigure}	 
	\caption{Clamp: Yield Strength (MPa)}
	\label{Fig:Clamp Nat Conv Yield}
\end{figure}

 \Cref{Eq:Energy,Eq:SDAS,Eq:Solid_fraction,Eq:YS,Eq:grain_growth_S,Eq:grain_growth_lambda,Eq:grain_growth_rate} are solved numerically using the software OpenCast \cite{shahane2019finite} with a finite volume method on a collocated grid. The variations of thermal conductivity, density and specific heat due to temperature are taken into account. Most practical die casting geometries are complex and require unstructured grids. In our work, we have first generated a tetrahedral mesh using GMSH \cite{geuzaine2009gmsh} and then divided into a hexahedral mesh using TETHEX \cite{Tethex_github}. The details of the numerical algorithm and verification and validation of OpenCast are discussed in previous publications \cite{shahane2019finite,shahane2019simulations}. A model geometry representing a clamp \cite{shahane2019finite} has been considered to illustrate the methodology. \Cref{Fig:Clamp} shows the clamp geometry with a mesh having 334000 hexahedral elements. It is important to assess the effects of natural convection. Hence, the clamp geometry is simulated for two cases viz. with and without natural convection. \Cref{Fig:Clamp Nat Conv Grain Size,Fig:Clamp Nat Conv Yield} plot grain size and yield strength contours with identical process conditions for both the cases. Since the solidification time is around 2.5 seconds, the velocities due to natural convection in the liquid metal are observed to be negligible. It is evident from the contours that there is no significant effect of natural convection and hence, it is neglected in all our further simulations.
\section{Optimization} \label{Sec:Optimization}
\begin{figure}[H]
	\centering
	\begin{subfigure}[t]{0.49\textwidth}
		\includegraphics[width=\textwidth]{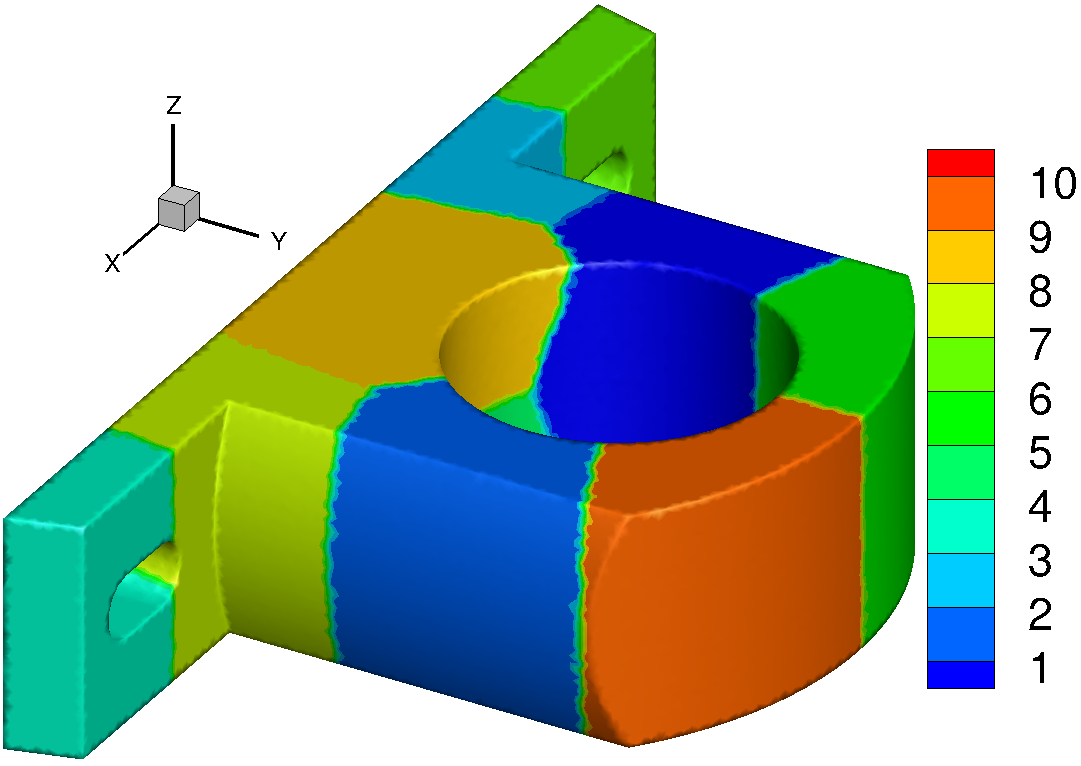}
		\caption{Domain Numbers}
		\label{Fig:Boundary Condition Representation by Domain Decomposition domain nos}
	\end{subfigure}
	\begin{subfigure}[t]{0.49\textwidth}
		\includegraphics[width=\textwidth]{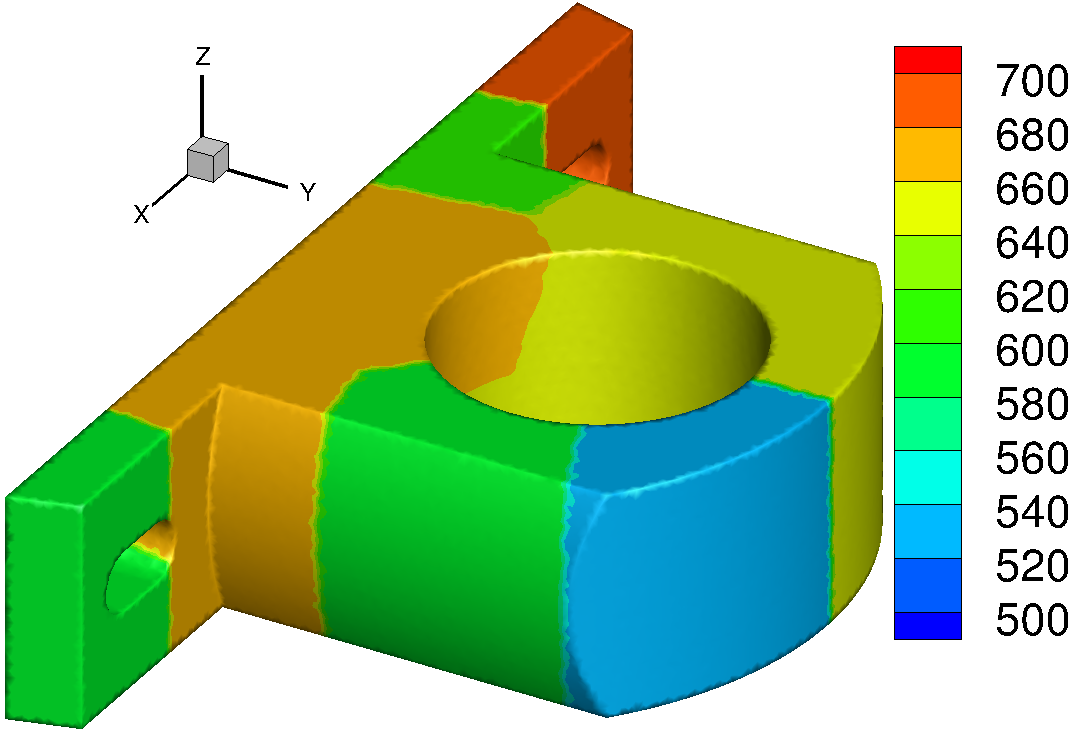}
		\caption{Randomly Assigned Values}
		\label{Fig:Boundary Condition Representation by Domain Decomposition sample}
	\end{subfigure}	 
	\caption{Domain Decomposition of the Boundary and Random Value Assignment}
	\label{Fig:Boundary Condition Representation by Domain Decomposition}
\end{figure}

In die casting the mold cavity is filled with molten metal and solidified. The heat is extracted from the cavity walls by flowing coolants (typically water) through the cooling lines made inside the die. The quality of the finished product depends on the rate of heat extraction which in turn depends on the temperature at the cavity walls. Due to complexity in the die geometry, the wall temperature varies locally. An optimal product quality can be achieved if the temperature distribution on the cavity walls and initial fill temperature are set properly. Thus, in this work, the following optimization problem with three objectives is proposed:
\begin{equation} 
\begin{aligned}
	&\text{Minimize } \{ f_1(T_{init}, \bm{T}_{wall}), f_2(T_{init}, \bm{T}_{wall}), f_3(T_{init}, \bm{T}_{wall}) \} \\
	&\text{subject to } 900 \leq T_{init} \leq 1100 \text{ K and } 500 \leq \bm{T}_{wall} \leq 700 \text{ K}
	\label{Eq:optim_problem}
\end{aligned}
\end{equation} 
where, $f_1=\text{solidification time}$, $f_2=\max{(\text{grain size})}$ and $f_3=-$min(yield strength). Minimizing the solidification time increases productivity. Reduction in grain size reduces susceptibility to cracking \cite{kimura2009effect} and improves mechanical properties of the product \cite{camicia2016grain}. Thus, minimization of the maximum value of grain size over the entire geometry is set as an optimization objective. Higher yield strength is desirable as it increases the elastic limit of the material. Hence, the minimum yield strength over the entire geometry is to be maximized. For convenience, this maximization problem is converted to minimization by multiplying by minus one. This explains the third objective function $f_3$. All the objectives are functions of the initial molten metal temperature ($T_{init}$) and mold wall temperature ($\bm{T}_{wall}$). The initial temperature is a single value in the interval $[900,1100]$ K which is higher than the liquidus temperature of the alloy. As discussed before, the mold wall temperature need not be uniform in die casting due to locally varying heat transfer to the cooling lines. Thus, in this work, the wall surface is decomposed into multiple domains with each domain having a uniform temperature boundary condition which is held constant with time during the entire solidification process. If the die design with cooling line placement and coolant flow conditions are included, the thermal analysis of the die can also be done to identify these domains. Due to the lack of this information, the wall is decomposed into ten domains using the KMeans classification algorithm from Scikit Learn \cite{scikit-learn}. \Cref{Fig:Boundary Condition Representation by Domain Decomposition domain nos} shows the domain decomposition with ten domain tags and \cref{Fig:Boundary Condition Representation by Domain Decomposition sample} shows a random sample of the boundary temperature with a single temperature value assigned uniformly to each domain. Thus, the input wall temperature ($\bm{T}_{wall}$) is a ten dimensional vector in the interval $[500,700]$ K which is lower than the solidus temperature of the alloy. Hence, this is a multi--objective optimization problem with three minimization objectives which are a function of eleven input temperatures.
\begin{figure}[H]
	\begin{subfigure}[t]{0.23\textwidth}
		\includegraphics[width=\textwidth]{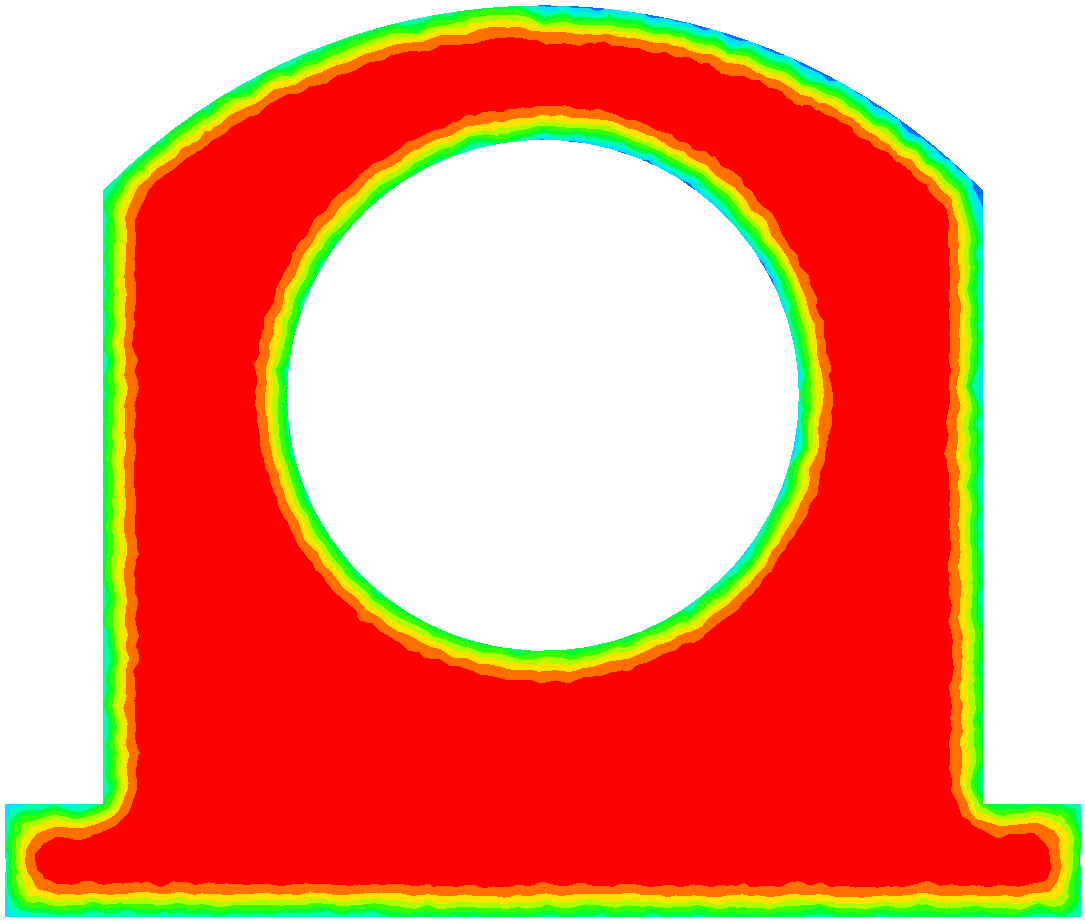}
		\caption{Time: 0.037 s}
	\end{subfigure}	
	\begin{subfigure}[t]{0.23\textwidth}
		\includegraphics[width=\textwidth]{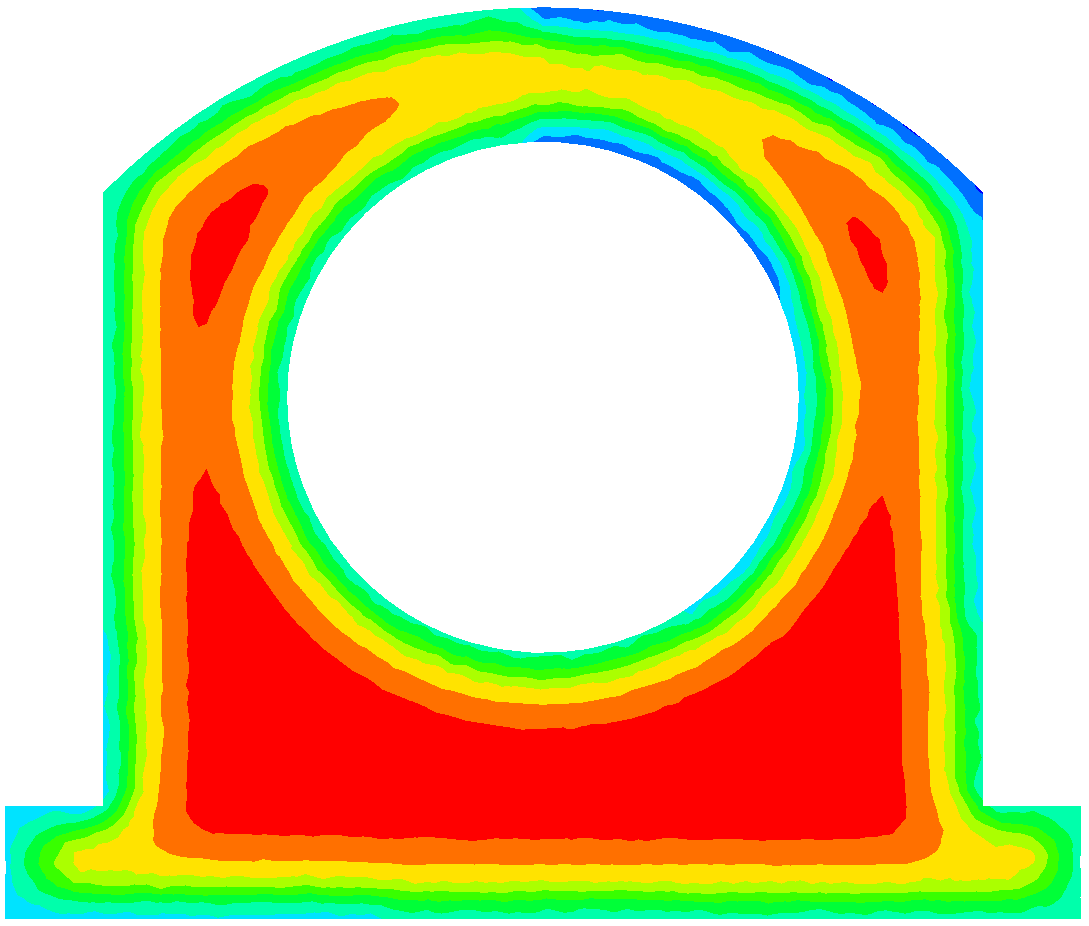}
		\caption{Time: 0.294 s}
	\end{subfigure}
	\begin{subfigure}[t]{0.23\textwidth}
		\includegraphics[width=\textwidth]{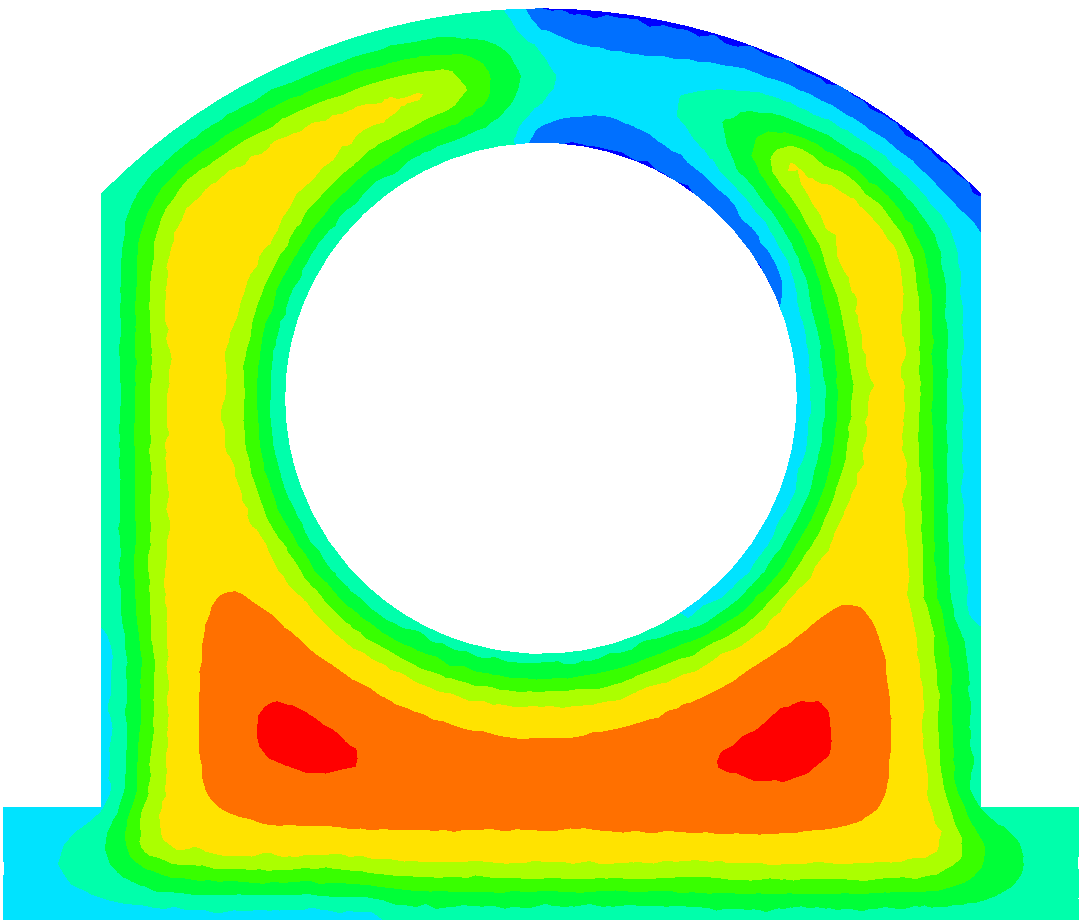}
		\caption{Time: 0.731 s}
	\end{subfigure}
	\begin{subfigure}[t]{0.23\textwidth}
		\includegraphics[width=\textwidth]{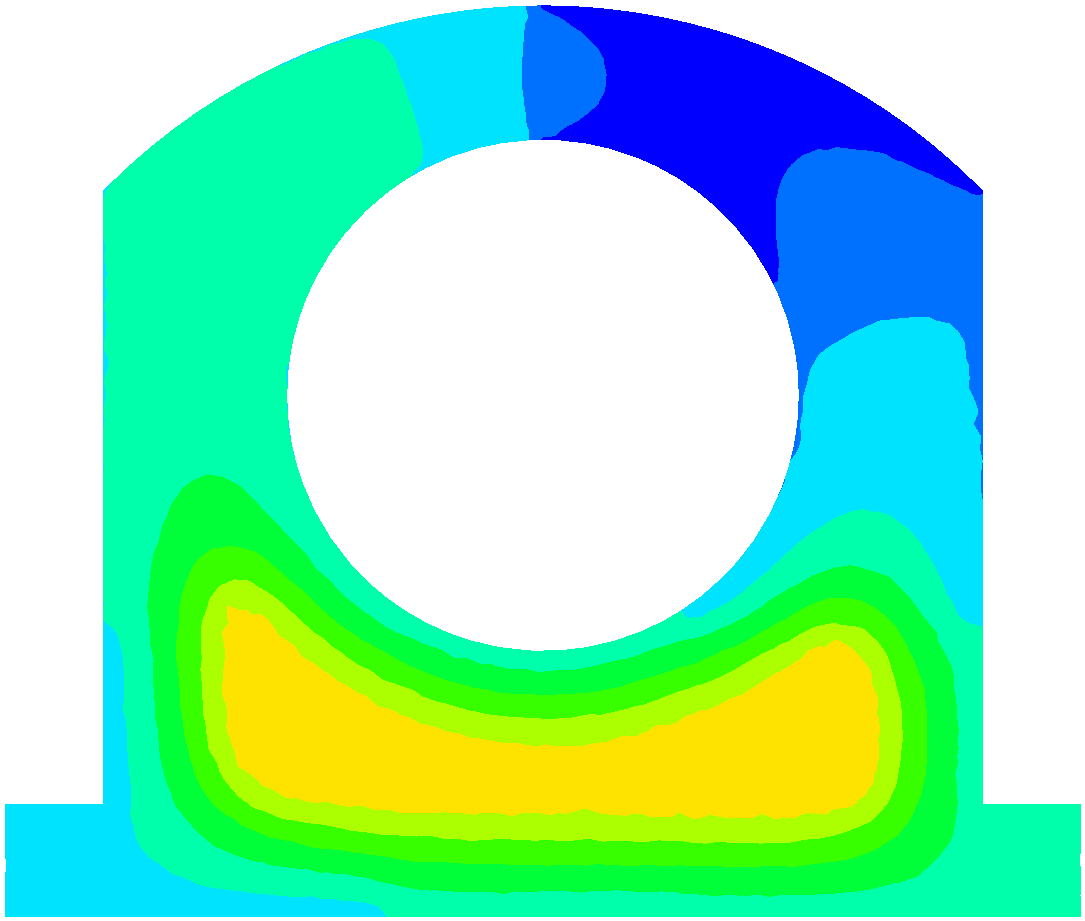}
		\caption{Time: 1.88 s}
	\end{subfigure}	
	\begin{subfigure}[t]{0.05\textwidth}
		\includegraphics[width=\textwidth]{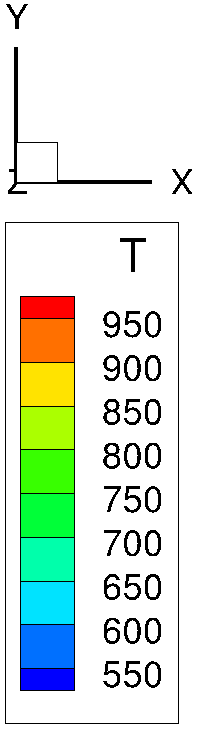}
	\end{subfigure}	
	\caption{Clamp Temperature Contours (K) (Solidification Time: 3.02 s)}
	\label{Fig:clamp_T}
\end{figure}
\begin{figure}[H]
	\begin{subfigure}[t]{0.23\textwidth}
		\includegraphics[width=\textwidth]{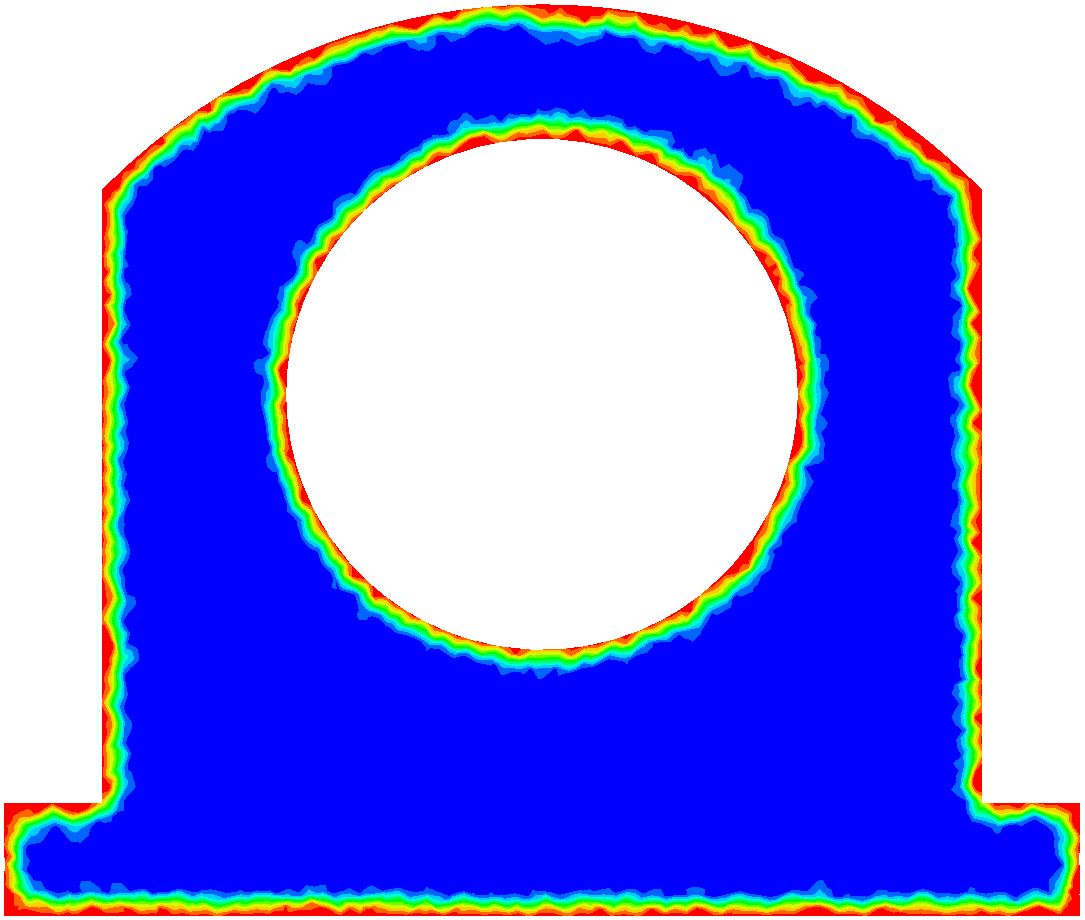}
		\caption{Time: 0.037 s}
	\end{subfigure}	
	\begin{subfigure}[t]{0.23\textwidth}
		\includegraphics[width=\textwidth]{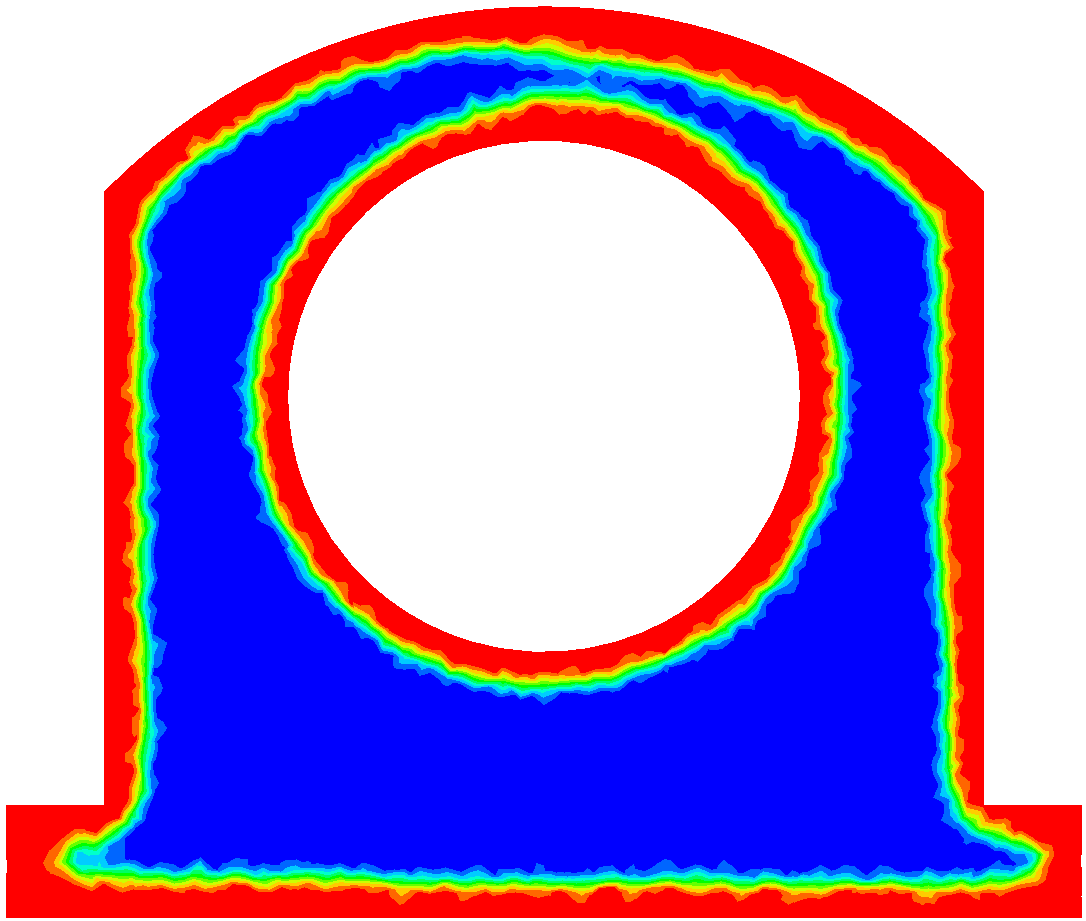}
		\caption{Time: 0.294 s}
	\end{subfigure}
	\begin{subfigure}[t]{0.23\textwidth}
		\includegraphics[width=\textwidth]{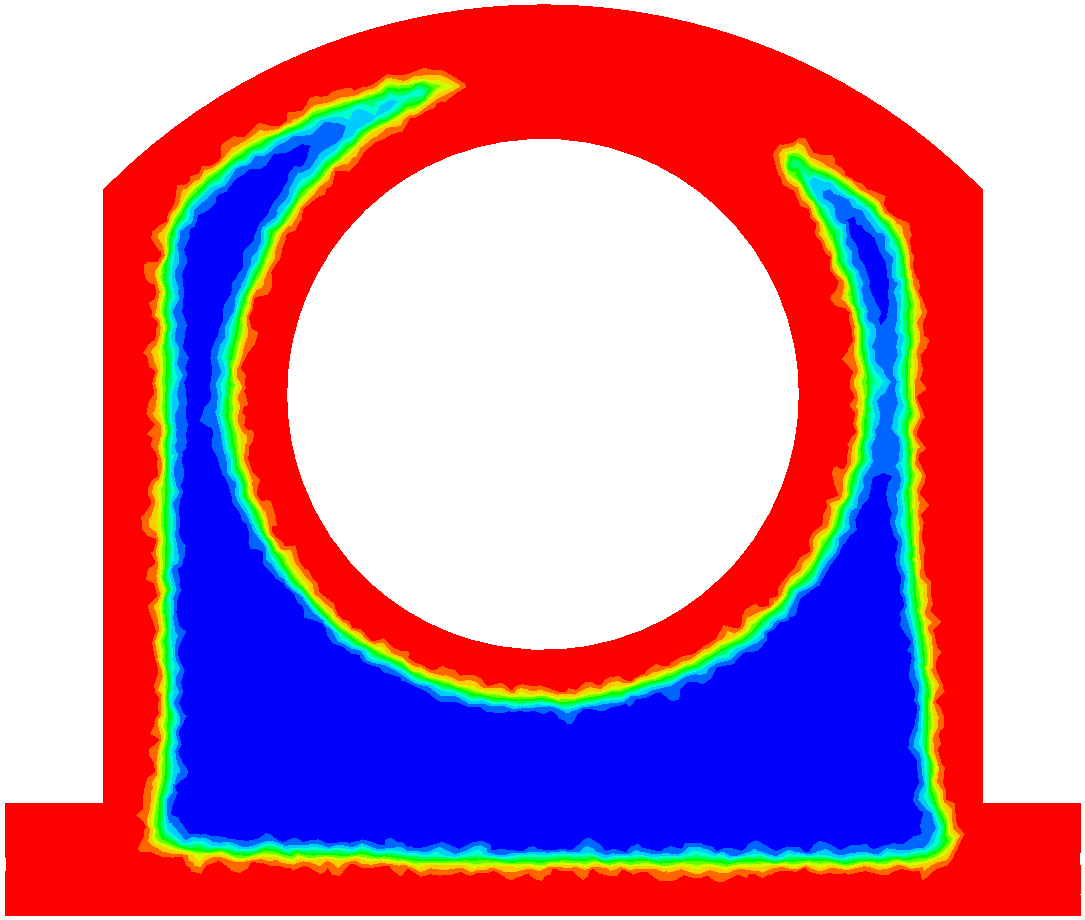}
		\caption{Time: 0.731 s}
	\end{subfigure}
	\begin{subfigure}[t]{0.23\textwidth}
		\includegraphics[width=\textwidth]{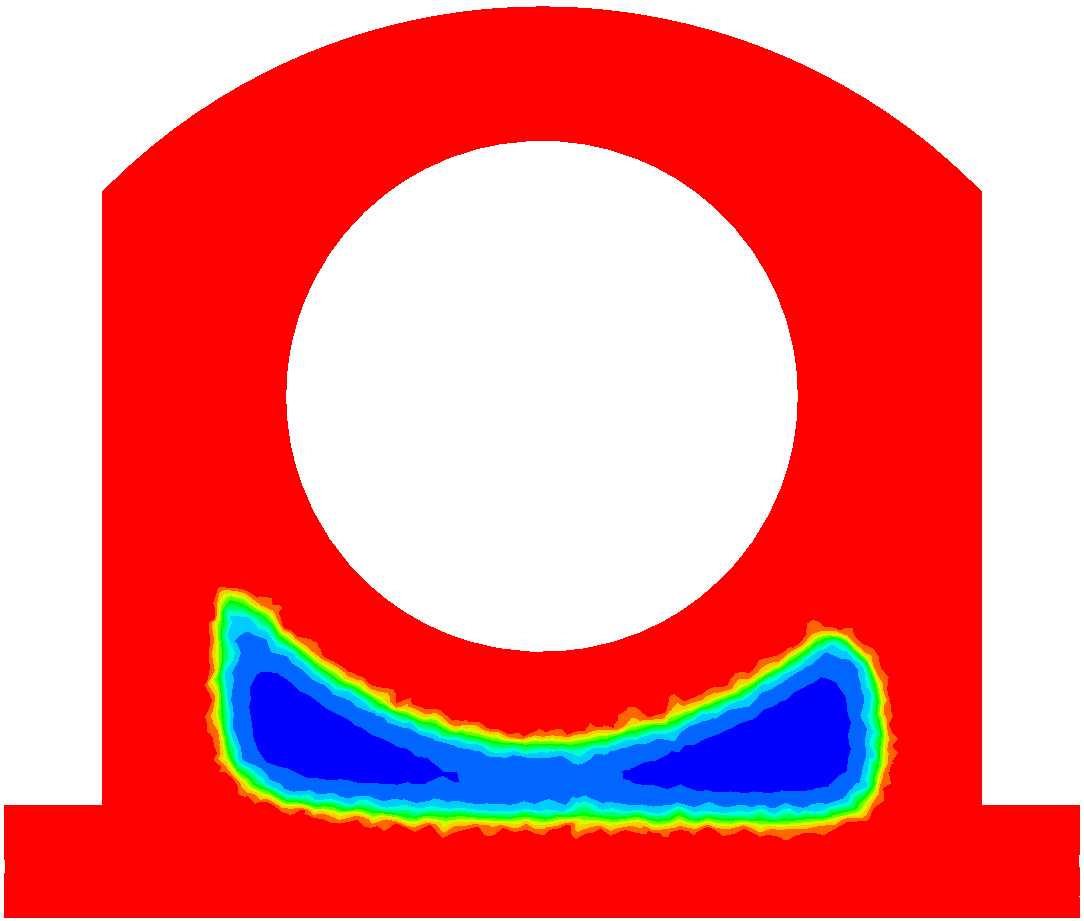}
		\caption{Time: 1.88 s}
	\end{subfigure}	
	\begin{subfigure}[t]{0.05\textwidth}
		\includegraphics[width=\textwidth]{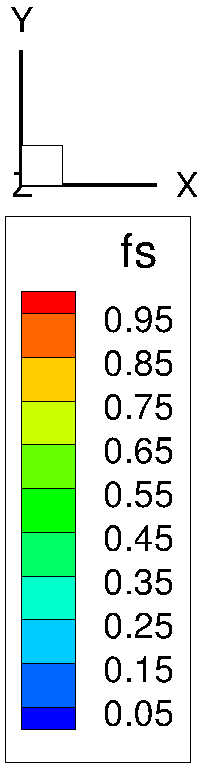}
	\end{subfigure}	
	\caption{Clamp Solid Fraction Contours (Solidification Time: 3.02 s)}
	\label{Fig:clamp_fs}
\end{figure}
\par
\begin{figure}[H]
	\begin{subfigure}[t]{0.49\textwidth}
		\includegraphics[width=\textwidth]{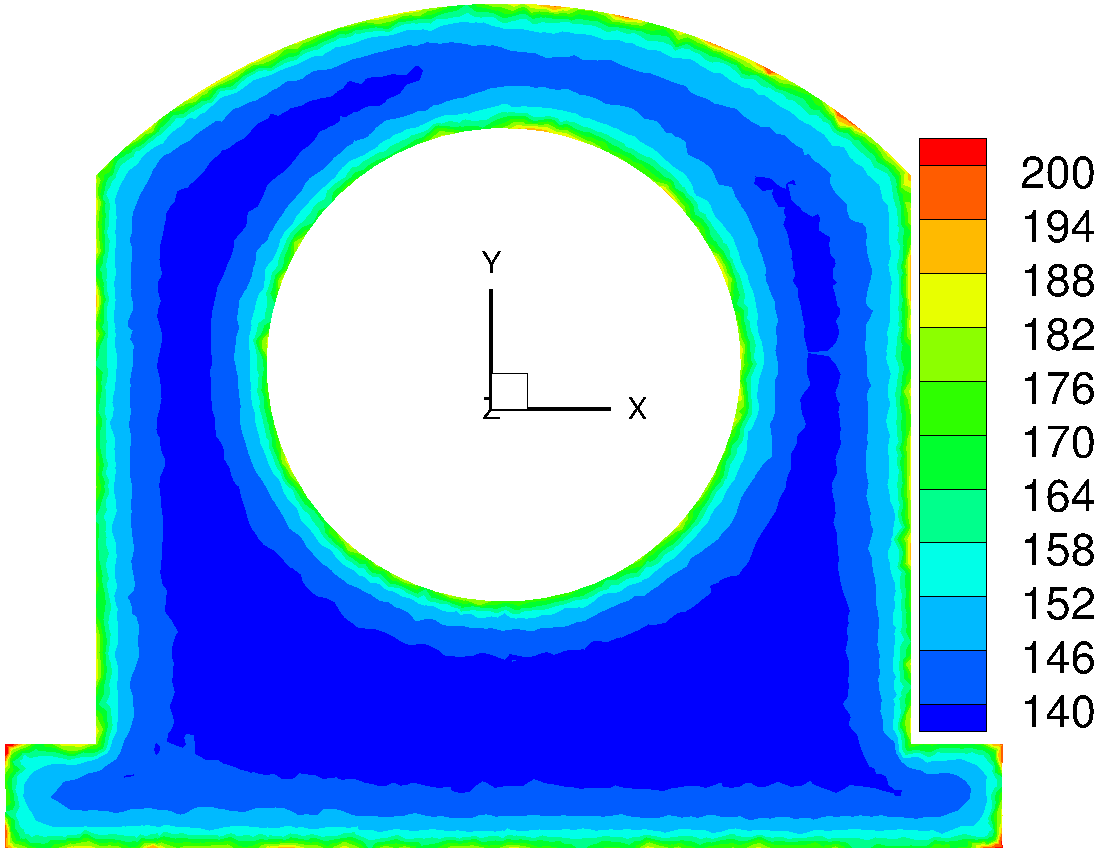}
		\caption{Yield Strength (MPa)}
		\label{Fig:clamp yield}
	\end{subfigure}
	\begin{subfigure}[t]{0.49\textwidth}
		\includegraphics[width=\textwidth]{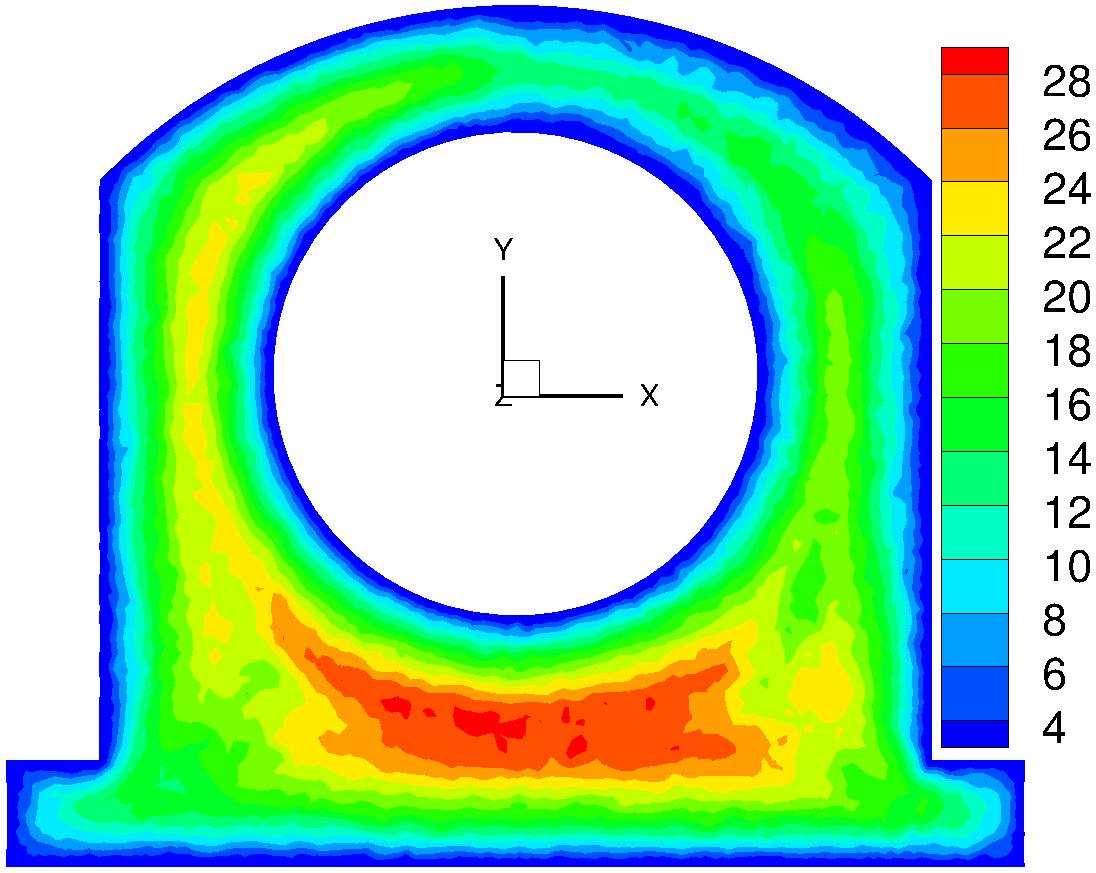}
		\caption{Grain Size ($\mu$m)}
		\label{Fig:clamp grain size}
	\end{subfigure}
	\caption{Clamp Microstructure Parameters}
	\label{Fig:clamp_microstruc}
\end{figure}

 \Cref{Fig:clamp_T,Fig:clamp_fs} plot temperature and solid fraction for different time steps during solidification for the sample shown in \cref{Fig:Boundary Condition Representation by Domain Decomposition sample} with $T_{init}=986$ K. It can be seen that the temperature and solid fraction contours are asymmetric due to non-uniform boundary temperature. For instance, domain number 10 is held at minimum temperature and thus, region near it solidifies first. \Cref{Fig:clamp_microstruc} plots final yield strength and grain size contours. The cooling rates and temperature gradients decrease with time as solidification progresses. Hence, the core regions which are thick take longer time to solidify. As the grains have more time to grow, the grain size is higher in the core region and correspondingly, the yield strength is lower. These trends along with the asymmetry are visible in \cref{Fig:clamp_microstruc}.
\subsection{Genetic Algorithm}
\subsubsection{Single Objective Optimization} \label{Sec:Single Objective Optimization}
Genetic algorithms (GAs) are global search algorithms based on the mechanics of natural selection and genetics. They apply the `survival of the fittest' concept on a set of artificial creatures characterized by strings. In the context of a GA, an encoded form of each input parameter is known as a gene. A complete set of genes which uniquely describe an individual (i.e., a feasible design) is known as a chromosome. The value of the objective function which is to be optimized is known as the fitness. The population of all the individuals at a given iteration is known as a generation. The overall steps in the algorithm are as follows:
\begin{enumerate}
	\item Initialize first generation with a random population
	\item Evaluate the fitness of the population
	\item Select parent pairs based on their fitness (better fitness implies higher probability of selection)
	\item Perform crossover to generate an offspring from each parent pair
	\item Mutate some genes randomly in the population
	\item Replace the current generation with the next generation 
	\item If termination condition is satisfied, return the best individual of the current generation; else go back to step 2
\end{enumerate}
\par There are multiple strategies discussed in the literature for each of the above steps \cite{goldberg1989genetic,coley1999introduction}. A brief overview of the methods used in this work is given here. The population is initialized using the Latin Hypercube Sampling strategy from the python package pyDOE \cite{pydoe}. Fitness evaluation is the estimation of the objective function which can be done by full scale computational model or by surrogate models. The number of objective function evaluations are typically of the order of millions and thus, step 2 becomes computationally most expensive step if a full scale model is used. Instead, it is common to use a surrogate model which is much cheaper to evaluate. In this work, a neural network based surrogate model is used, details of which are provided in \cref{Sec:Neural Network}. Tournament selection is used to choose the parent pairs to perform crossover. The idea is to choose four individuals at random and select two out of them which have better fitness. Note that since the optimization is cast as a minimization problem, lower fitness value is desired. Uniform crossover is used to recombine the genes of the parents to generate the offspring with a crossover probability of 0.9. Random mutation of the genes of the entire generation is performed with a mutation probability of 0.1. Thereafter, the old generation is replaced by this new generation. Note that the elitist version of GA is used which passes down the fittest individual of the previous generation to this generation as it is. Elitism was found helpful as it ensures that the next generation is at least as good as the previous generation.
\subsubsection{Multi--Objective Optimization}
The simultaneous optimization of multiple objectives is different than the single objective optimization problem. In a single objective problem, the best design which is usually the global optimum (minimum or maximum) is searched for. On the other hand, for multi--objective problem, there may not exist a single optimum which is the best design or global optimum with respect to all the objectives simultaneously. This happens due to the conflicting nature of objectives i.e., improvement in one can cause deterioration of the other objectives. Thus, typically there is a set of Pareto optimal solutions which are superior to rest of the solutions in the design space which are known as dominated solutions. All the Pareto optimal solutions are equally good and none of them can be prioritized in the absence of further information. Thus, it is useful to have a knowledge of multiple non-dominated or Pareto optimal solutions so that a single solution can be chosen out of them considering other problem parameters.
\par One possible way of dealing with multiple objectives is to define a single objective as a weighted sum of all the objectives. Any single objective optimization algorithm can be used to obtain an optimal solution. Then the weight vector is varied to get a different optimal solution. The problem with this method is that the solution is sensitive to the weight vector and choosing the weights to get multiple Pareto optimal solutions is difficult for a practical engineering problem. Multi--objective GAs attempt to handle all the objectives simultaneously and thus, annihilating the need of choosing the weight vector. \citet{konak2006multi} have discussed various popular multi--objective GAs with their benefits and drawbacks. In this work, the Non-dominated Sorting Genetic Algorithm II (NSGA--II) \cite{deb2002fast} which is a fast and elitist version of the NSGA algorithm \cite{srinivas1994muiltiobjective} is used. The NSGA--II algorithm to march from a given generation of population size $N$ to a next generation of same size is as follows:
\begin{enumerate}
	\item Select parent pairs based on their rank computed before (lower rank implies higher probability of selection)
	\item Perform crossover to generate an offspring from each parent pair
	\item Mutate some genes randomly in the population thus forming the offspring population
	\item Merge the parent and offspring population thus giving a set of size $2 \times N$
	\item Evaluate the fitness of the population corresponding to each objective
	\item Divide the population into multiple non-dominated levels also known as fronts
	\item Compute the crowding distance for each individual along each front
	\item Sort the population based on front number and crowding distances and rank them
	\item Choose the best set of $N$ individuals as next generation (i.e., $N$ individuals with lowest ranks)
	\item If termination condition is satisfied, return the best front of the current generation as an approximation of the Pareto optimal solutions; else go back to step 1
\end{enumerate}
Before iterating over the above steps, some pre-processing is required. A random population of size $N$ is initiated and steps 5--8 are implemented to rank the initial generation. The parent selection, crossover and mutation steps are identical to the single objective GA described in \cref{Sec:Single Objective Optimization}. The algorithms for remainder of the steps can be found in the paper by \citet{deb2002fast}. Ranking the population by front levels and crowding distance enforces both elitism and diversity in the next generation.
\subsection{Neural Network} \label{Sec:Neural Network}
The fitness evaluation step of the GA requires a way to estimate the outputs corresponding to the set of given inputs. Typically, the number of generations can be of the order of thousands with several hundreds of population size per generation and thus, the total number of function evaluations can be around hundred thousands or more. It is computationally difficult to run the full scale numerical estimation software. Thus, a surrogate model is trained which is cheap to evaluate. A separate neural network is trained for each of the three optimization objectives (\cref{Eq:optim_problem}). \citet{hornik1989multilayer} showed that under mild assumptions on the function to be approximated, a neural network can achieve any desired level of accuracy by tuning the hyper-parameters. The building block of a neural network is known as a neuron which has multiple inputs and gives out single output by performing the following operations:
\begin{enumerate}
	\item Linear transformation: $a=\sum_{i=1}^{n}{w_i x_i } + b$; where, $\{x_1,x_2,...,x_n\}$ are $n$ scalar inputs, $w_i$ are the weights and $b$ is a bias term
	\item Nonlinear transformation applied element-wise: $y=\sigma(a)$; where, $y$ is a scalar output and $\sigma$ is the activation function
\end{enumerate} 
Multiple neurons are stacked in a layer and multiple layers are connected together to form a neural network. The first and last layers are known as input and output layers respectively. Information flows from the input to output layer through the intermediate layers known as hidden layers. Each hidden layer adds nonlinearity to the network and thus, a more complex function can be approximated successfully. At the same time, having large number of neurons can cause high variance and thus, blindly increasing the depth of the network may not always help. The number of neurons in the input and output layers is defined by the problem specification. On the other hand, number of hidden layers and neurons has to be fine tuned to have low bias and low variance. The interpolation error of the network is quantified as a loss function which is a function of the weights and bias. 
\begin{table}[H]
	\centering
	\resizebox{\textwidth}{!}{%
		\begin{tabular}{|c|c|c|c|c|c|c|}
			\hline
			\begin{tabular}[c]{@{}c@{}}Name of\\Network\end{tabular} & \begin{tabular}[c]{@{}c@{}}No. of\\Hidden Layers\end{tabular} & \begin{tabular}[c]{@{}c@{}}No. of Neurons\\per Hidden Layer\end{tabular} & Learning Rate & \begin{tabular}[c]{@{}c@{}}No. of \\Epochs\end{tabular} & L2 $\lambda$ & \begin{tabular}[c]{@{}c@{}}Dropout\\Factor\end{tabular} \\ \hline
			Sol. Time & 4 & 50 & 0.001 & 300 & 0.004 & 0 \\ \hline
			Max. Grain & 6 & 75 & 0.001 & 300 & 0.005 & 0.2 \\ \hline
			Min. Yield & 4 & 25 & 0.003 & 400 & 0.01 & 0 \\ \hline
		\end{tabular}%
	}
	\caption{Neural Network Hyper--Parameters}
	\label{Tab:Neural_Network_Parameters}
\end{table}
\begin{table}[H]
	\centering
		\begin{tabular}{|c|c|c|c|}
			\hline
			Name of Network & Training Error & Tesing Error & Validation Error \\ \hline
			Sol. Time & 0.90\% & 1.01\% & 1.08\% \\ \hline
			Max. Grain & 1.29\% & 2.04\% & 1.92\% \\ \hline
			Min. Yield & 0.25\% & 0.38\% & 0.43\% \\ \hline
		\end{tabular}%
	\caption{Average Percent Training, Testing and Validation Errors}
	\label{Tab:Neural_Network_Errors}
\end{table}
\begin{figure}[H]
	\centering
	\begin{subfigure}[t]{0.32\textwidth}
		\includegraphics[width=\textwidth]{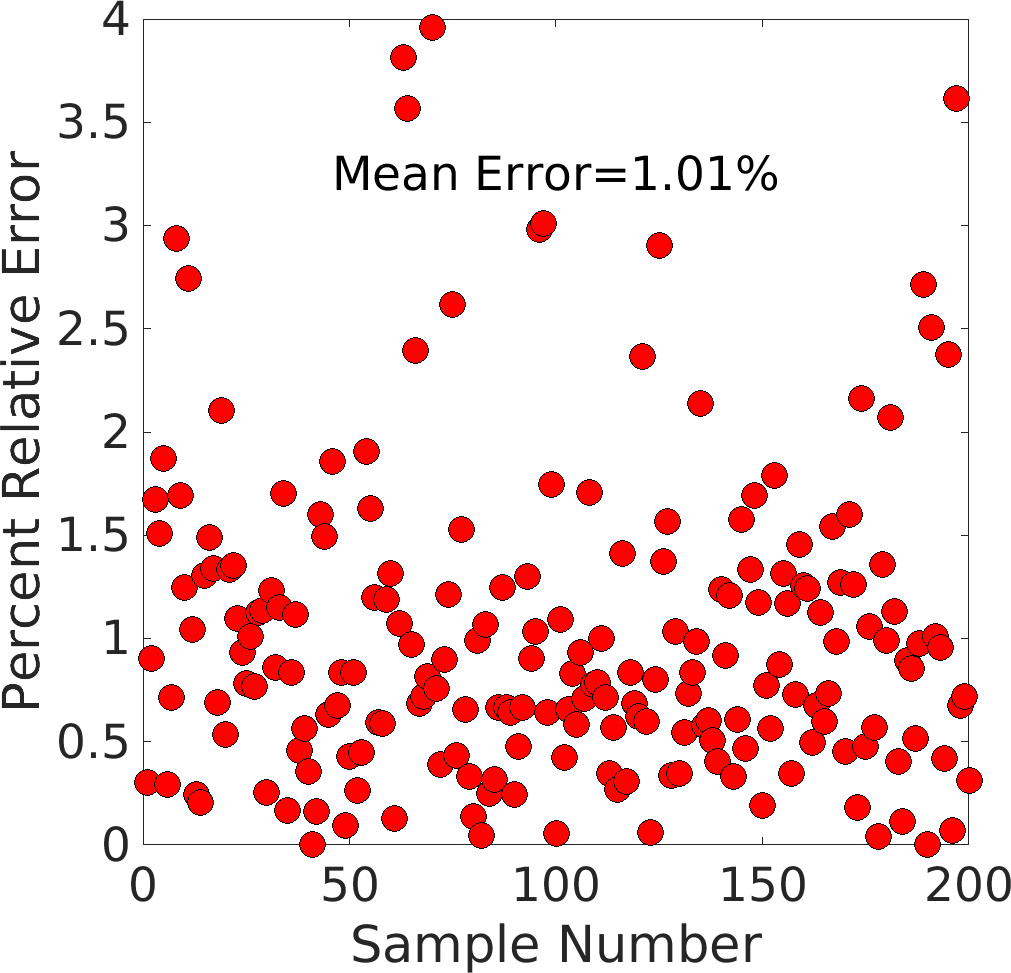}
		\caption{Solidification Time}
	\end{subfigure}
	\begin{subfigure}[t]{0.32\textwidth}
		\includegraphics[width=\textwidth]{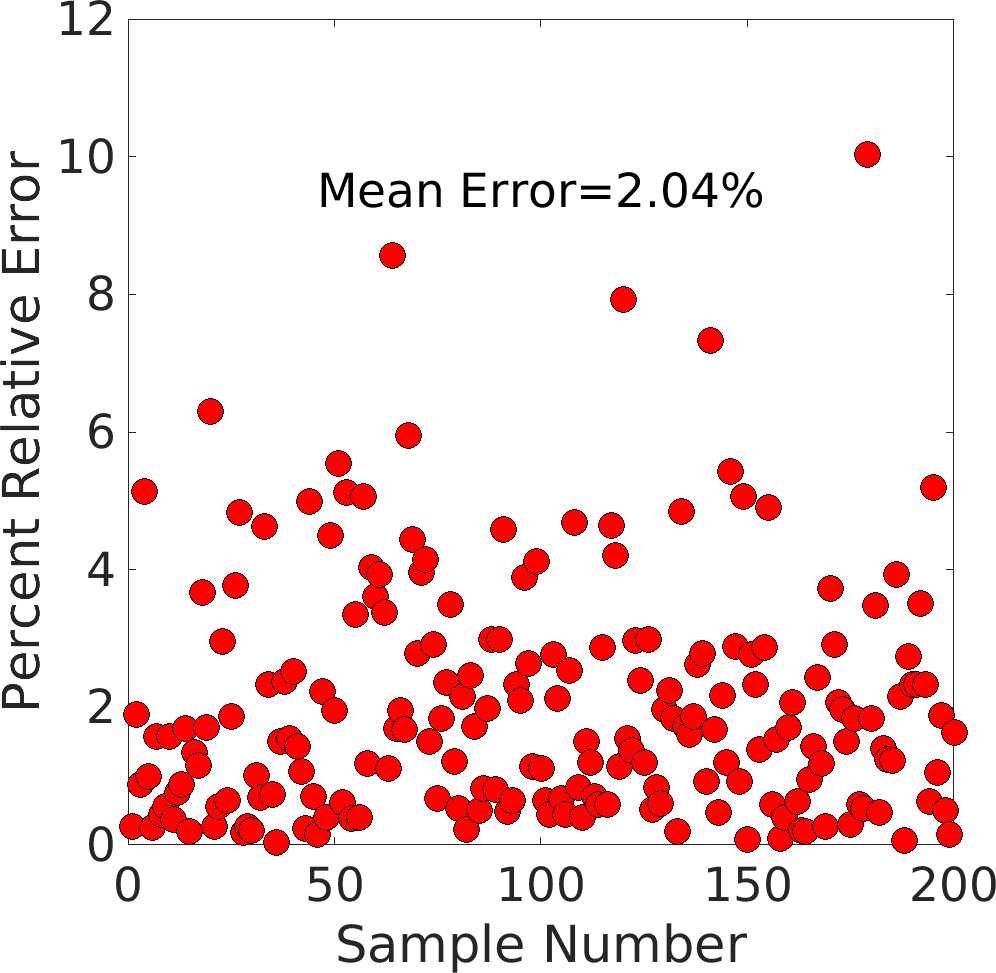}
		\caption{Max. Grain Size}
	\end{subfigure}	 
	\begin{subfigure}[t]{0.32\textwidth}
		\includegraphics[width=\textwidth]{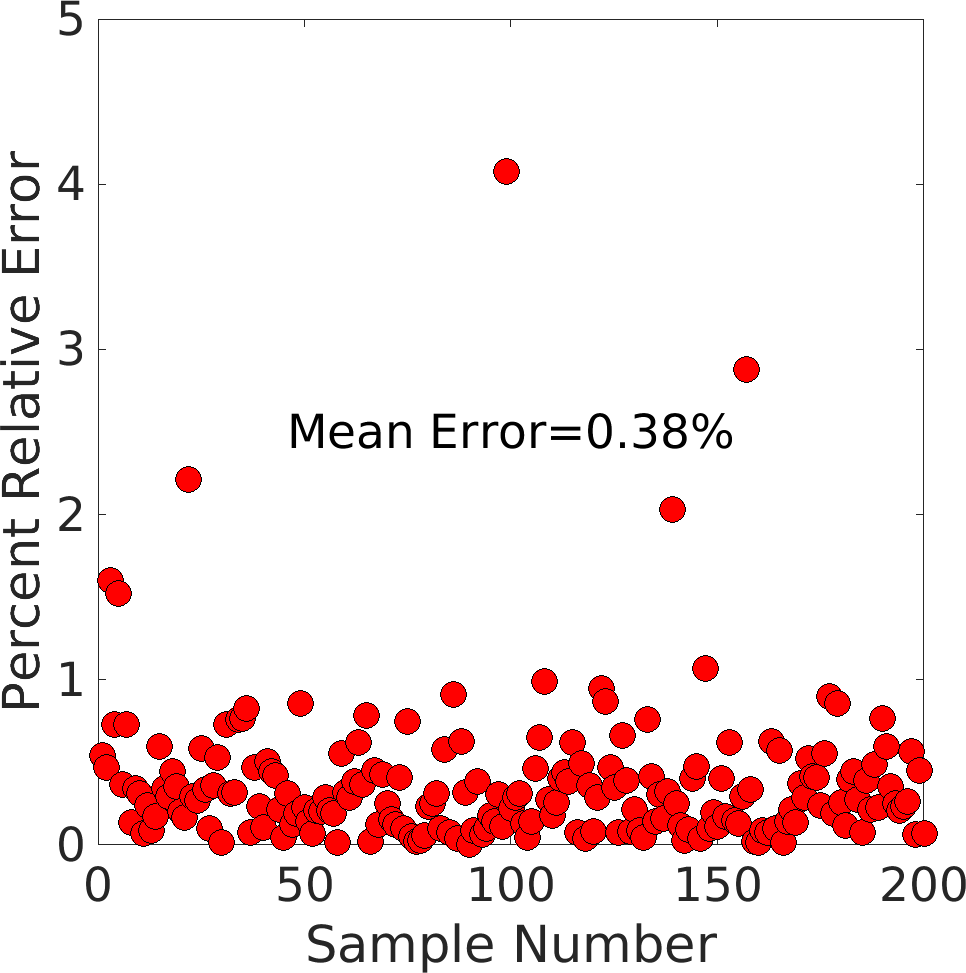}
		\caption{Min. Yield Str.}
	\end{subfigure}	 
	\caption{Neural Networks: Error Estimates for 200 Testing Samples}
	\label{Fig:Neural Networks: Error Estimates}
\end{figure}
\par A numerical optimization algorithm coupled with gradient estimation by the backpropagation algorithm \cite{chauvin2013backpropagation} is used to estimate the optimal weights and bias which minimize the loss function for a given training set. This is known as training process. An approach described by \citet{goodfellow2016deep} is used to select the hyper--parameters and train the neural network. Mean square error is set as the loss function. A set of 500 random samples is used for training and two different sets of 200 each are used for validation and testing. The number of inputs and outputs of the problem specify the number of neurons in the input and output layers respectively. Here, there are three neural networks with 11 input neurons and one output neuron each (11 temperatures and three objectives mentioned in \cref{Eq:optim_problem}). The number of hidden layers and hidden neurons, learning rate, optimizer, number of epochs and regularization constant are the hyper--parameters which are problem specific. They are varied in a range and then chosen so that both the training and validation errors are minimized simultaneously. The accuracy of prediction is further checked on an unseen test data. This overall approach is used for training the neural network with low bias and low variance. In this work, the neural network is implemented using the Python library Tensorflow \cite{tensorflow2015-whitepaper} with a high level API Keras \cite{chollet2015keras}. After testing various optimizers available in Keras, it is found that the Adam optimizer is the most suitable here. The parameters $\beta_1=0.9$ and $\beta_2=0.999$ are set as suggested in the original paper by \citet{kingma2014adam} and the `amsgrad' option is switched on. The activation functions used for all the hidden and output layers are ReLU and identity respectively. This choice of activation functions is generally recommended for regression problems \cite{goodfellow2016deep}. A combination of L2 regularization with dropout is used to control the variance. The training is stopped when the loss function changes minimally after a particular epoch. This is known as the stopping criteria. After varying the hyper-parameters in a range, it is found that for the values listed in \cref{Tab:Neural_Network_Parameters}, the training, testing and validation errors (\cref{Tab:Neural_Network_Errors}) are simultaneously minimized. Moreover, since all the three errors are low and close to each other, it shows that the bias and variance are low. \Cref{Fig:Neural Networks: Error Estimates} plots percent relative error for 200 testing samples. It can be seen that all the three neural networks are able to predict with acceptable accuracy.
\section{Assessment of Genetic Algorithm on Simpler Problems} \label{Sec:Assessment of Genetic Algorithm on Simpler Problems}
It is difficult to visualize the variation of an output with respect to each of the eleven inputs. Hence in this section, two problems are considered which are simplified versions of the actual problem. In the first case, the boundary temperature is set to a uniform value and hence, there are only two scalar inputs: $(T_{init}, T_{wall})$. For the second case, the initial temperature is held constant ($T_{init}=1000$ K) and the boundary is split into two domains instead of ten. Thus, again there are two scalar inputs: $(T_{wall}^{(1)}, T_{wall}^{(2)})$. $T_{wall}^{(1)}$ is assigned to domain numbers 1--5 and $T_{wall}^{(2)}$ to domain numbers 6--10 (\cref{Fig:Boundary Condition Representation by Domain Decomposition domain nos}). The ranges of wall and initial temperatures are the same as before (\cref{Sec:Optimization}). Such a simplified analysis gives an insight into the actual problem. Moreover, since these are problems with two inputs, the optimization can be performed by brute force parameter sweep and compared to the genetic algorithm. This helps to fine tune the parameters and assess the accuracy of the GA.
\subsection{Single Objective Optimization} 
\begin{figure}[H]
	\centering
	\begin{subfigure}[t]{0.49\textwidth}
		\includegraphics[width=\textwidth]{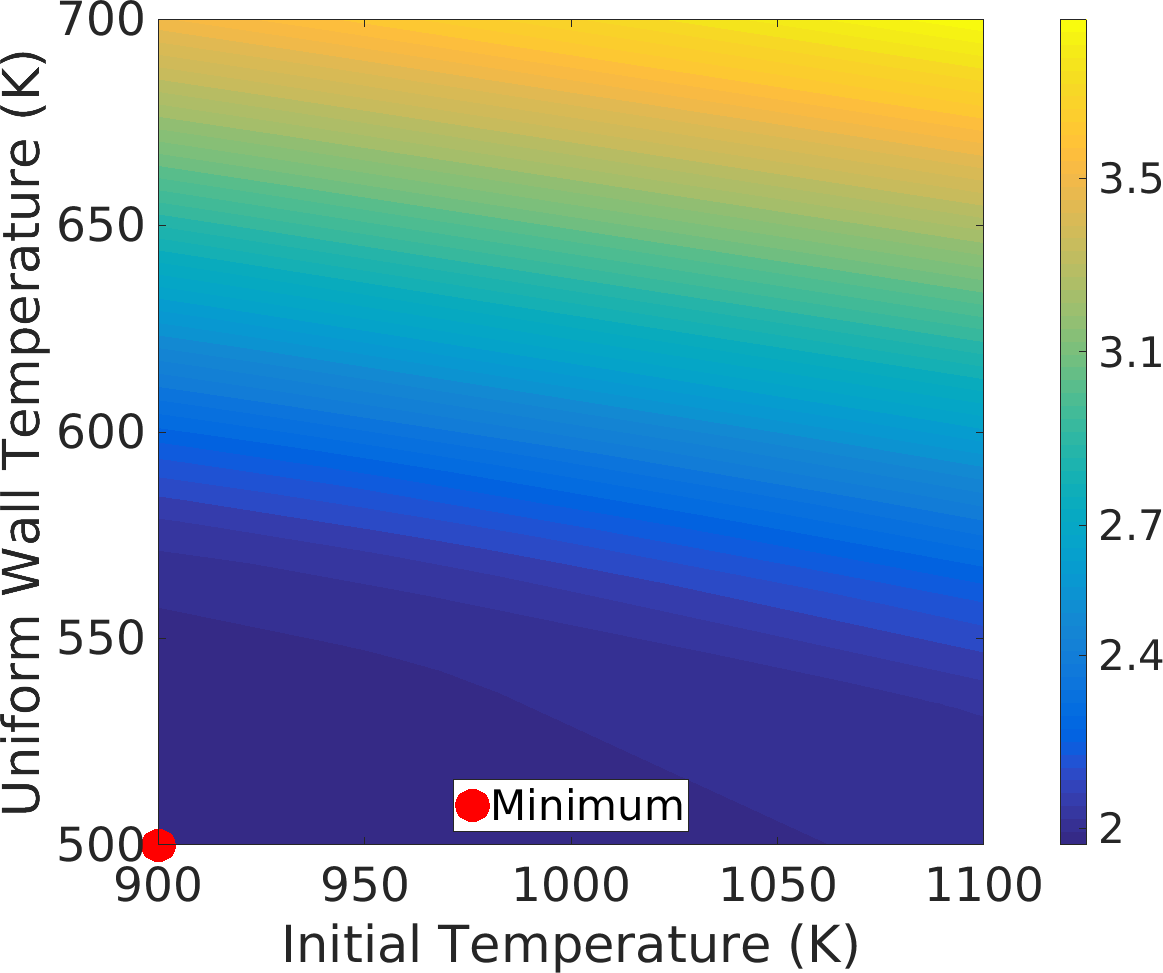}
		\caption{Solidification Time (s)}
		\label{Fig:Parameter Sweep: Uniform Boundary Temperature Sol Time}
	\end{subfigure}
	\begin{subfigure}[t]{0.49\textwidth}
		\includegraphics[width=\textwidth]{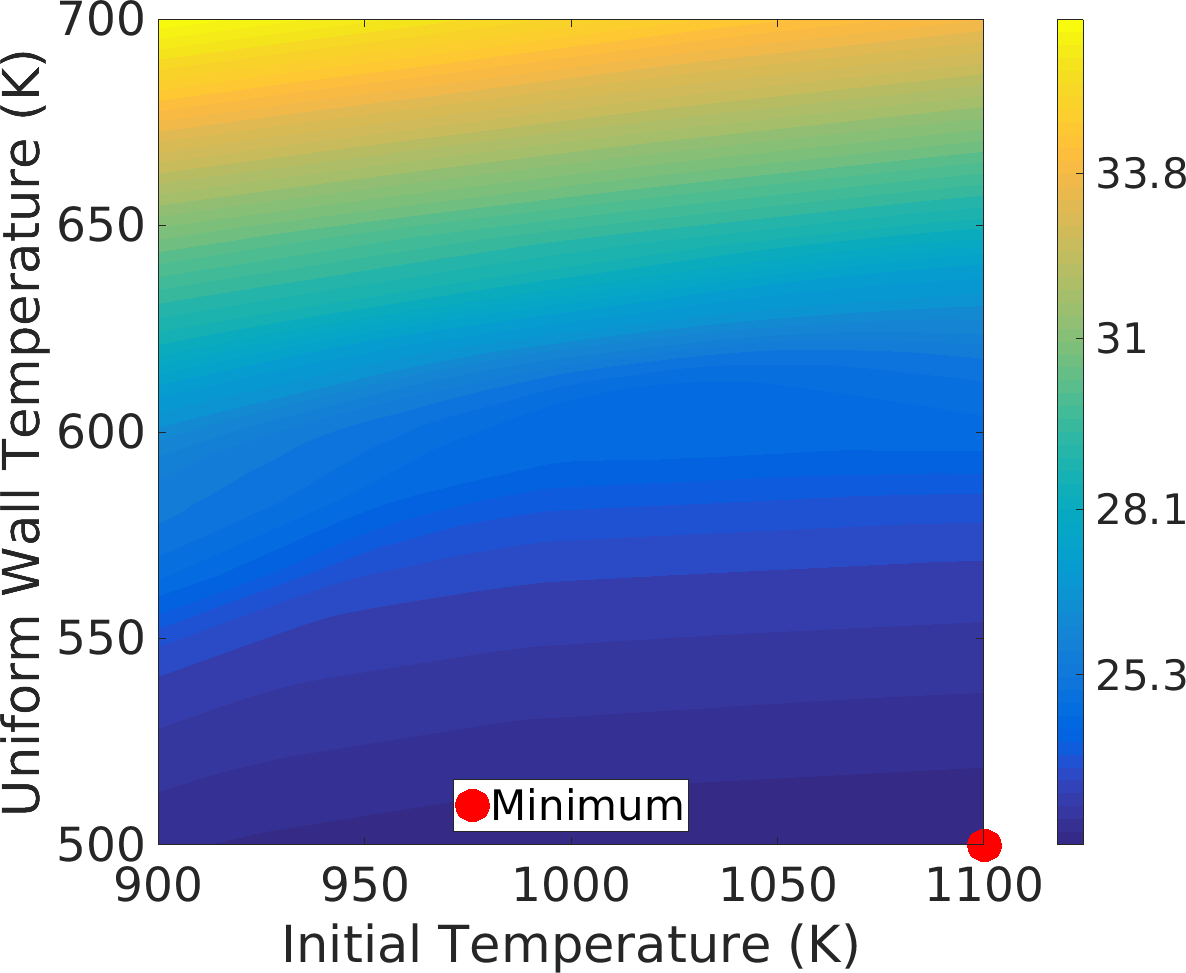}
		\caption{Max. Grain Size ($\mu$m)}
		\label{Fig:Parameter Sweep: Uniform Boundary Temperature Max Grain}
	\end{subfigure}	 
	\begin{subfigure}[t]{0.49\textwidth}
		\includegraphics[width=\textwidth]{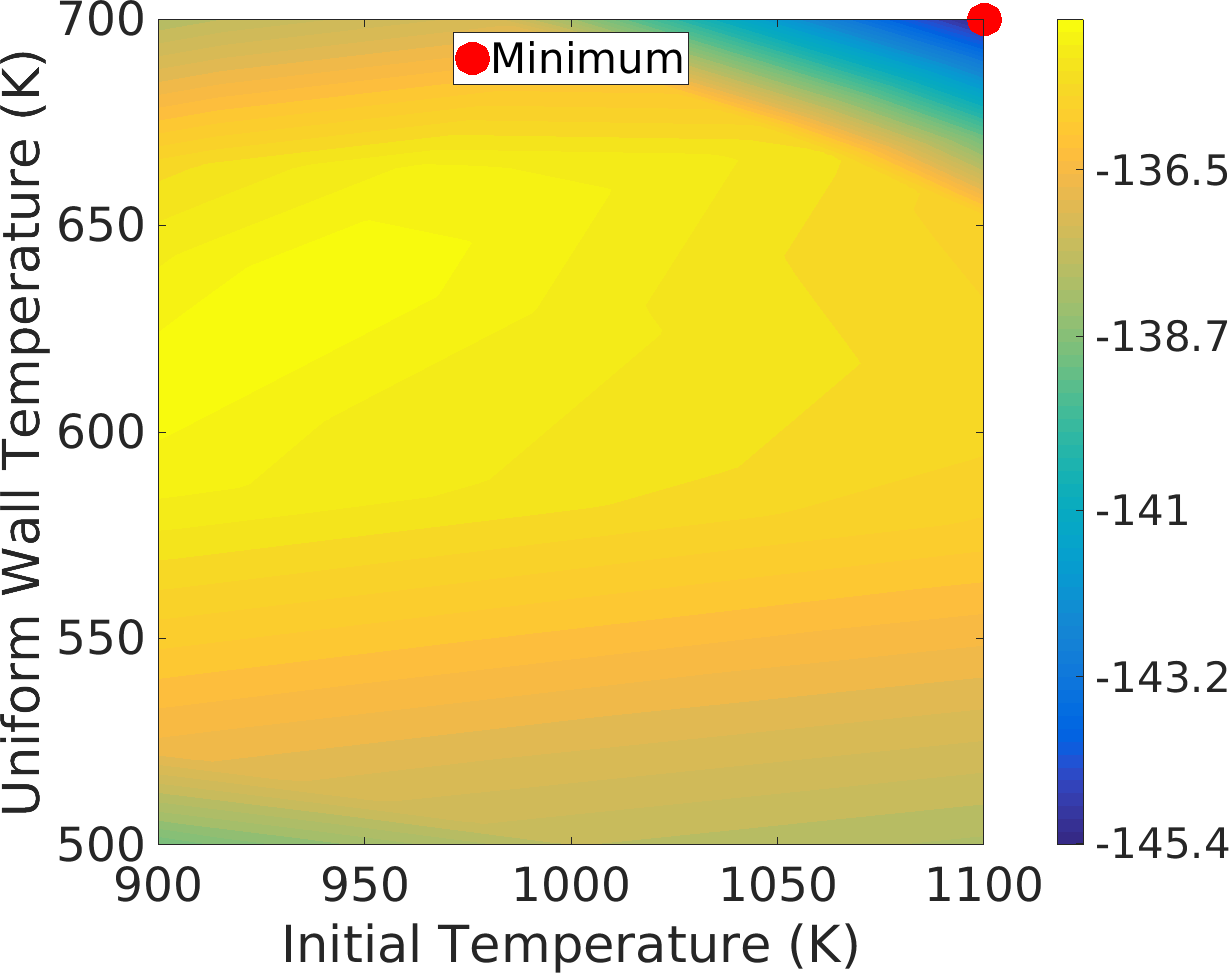}
		\caption{Min. Yield Str. (MPa)}
		\label{Fig:Parameter Sweep: Uniform Boundary Temperature Min Yield}
	\end{subfigure}	 
	\caption{Parameter Sweep: Uniform Boundary Temperature}
	\label{Fig:Parameter Sweep: Uniform Boundary Temperature}
\end{figure}

\begin{figure}[H]
	\centering
	\begin{subfigure}[t]{0.49\textwidth}
		\includegraphics[width=\textwidth]{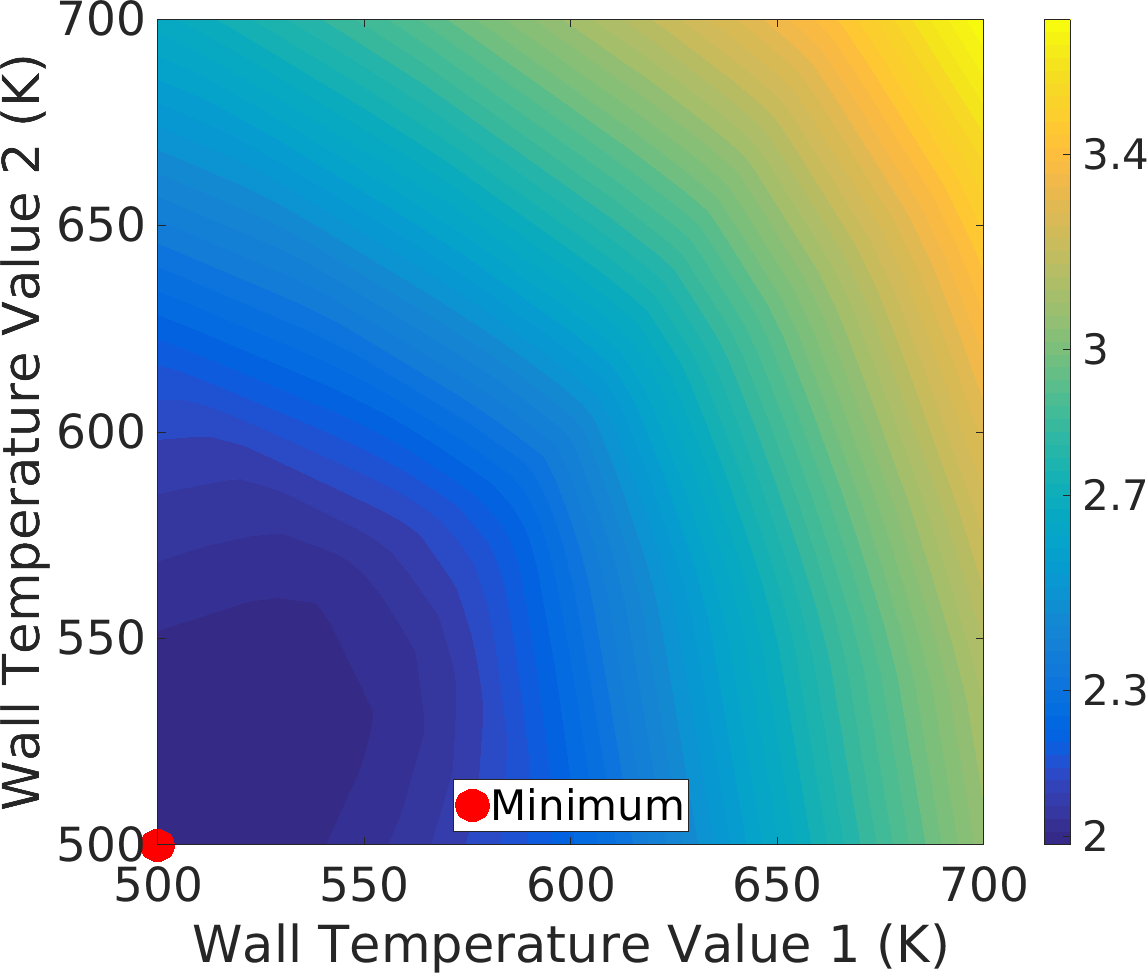}
		\caption{Solidification Time (s)}
		\label{Fig:Parameter Sweep: Split Boundary Temperature Sol Time}
	\end{subfigure}
	\begin{subfigure}[t]{0.49\textwidth}
		\includegraphics[width=\textwidth]{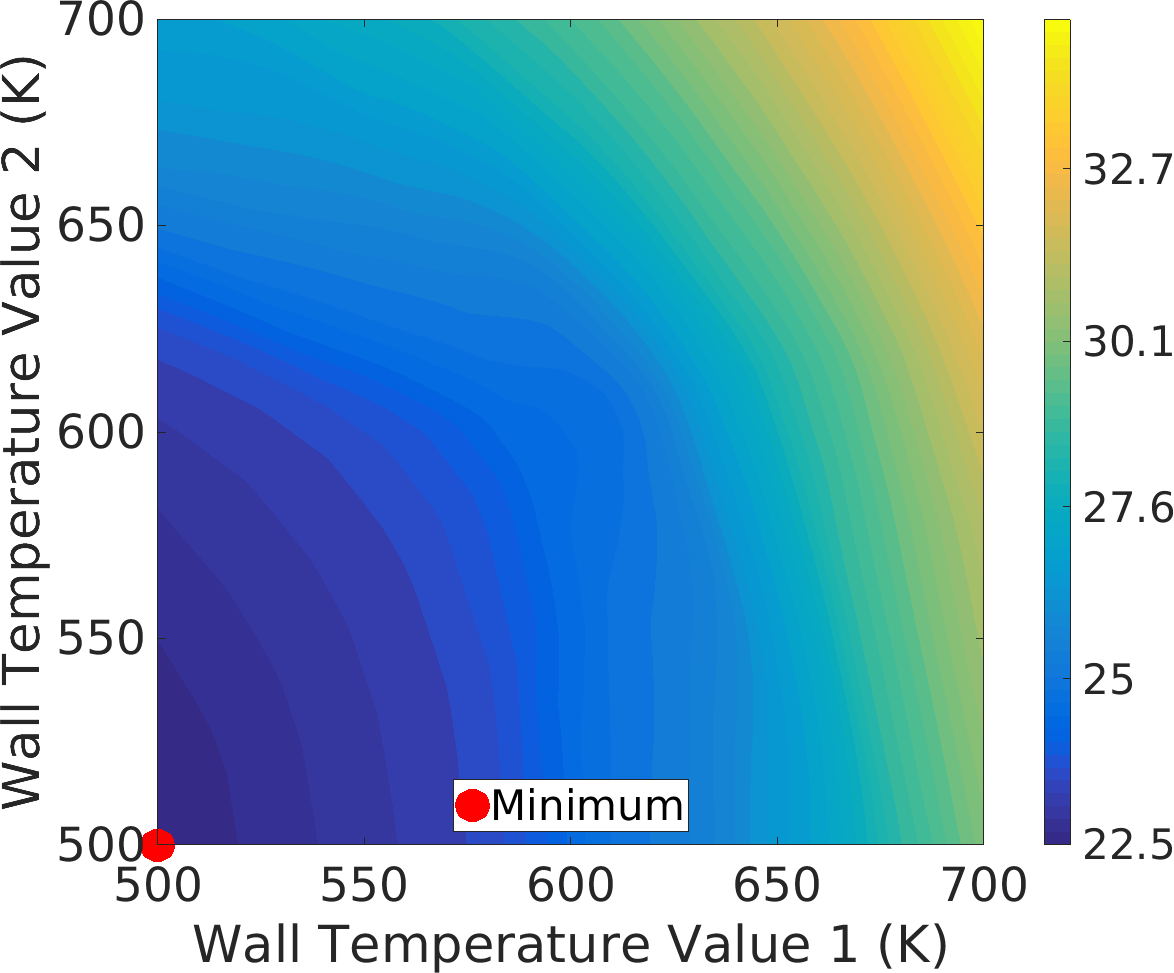}
		\caption{Max. Grain Size ($\mu$m)}
		\label{Fig:Parameter Sweep: Split Boundary Temperature Max Grain}
	\end{subfigure}	 
	\begin{subfigure}[t]{0.49\textwidth}
		\includegraphics[width=\textwidth]{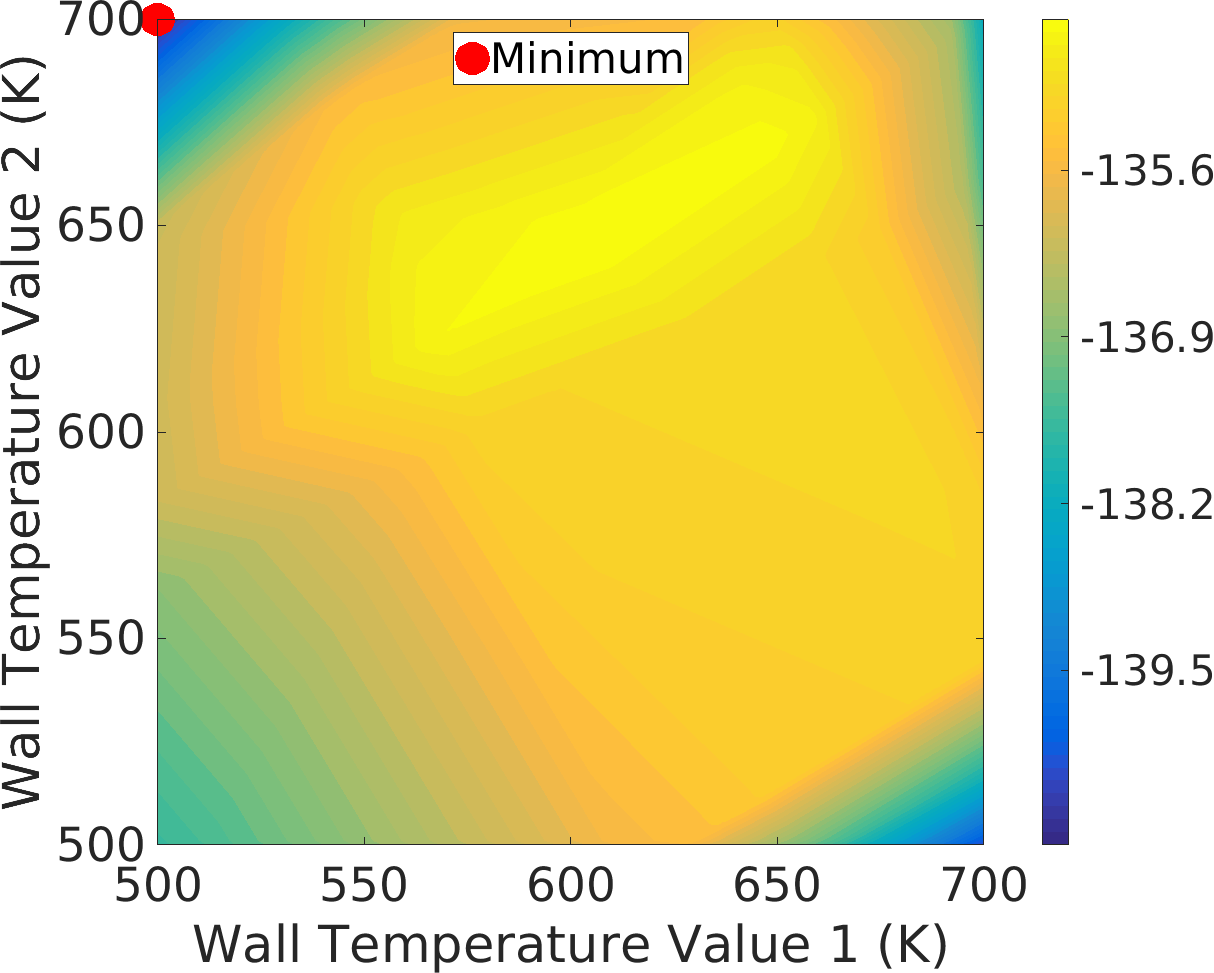}
		\caption{Min. Yield Str. (MPa)}
		\label{Fig:Parameter Sweep: Split Boundary Temperature Min Yield}
	\end{subfigure}	 
	\caption{Parameter Sweep: Split Boundary Temperature with $T_{init}=1000$ K}
	\label{Fig:Parameter Sweep: Split Boundary Temperature}
\end{figure}

In this section, all the objectives are analyzed individually. A two dimensional mesh of size 40,000 with 200 points in each input dimension is used for this analysis. The outputs are estimated from the neural networks for each of these points. \Cref{Fig:Parameter Sweep: Uniform Boundary Temperature} plots the response surface contours for each of the three objectives with their corresponding minima. The minima are estimated from the 40,000 points. The $X$ and $Y$ axes are initial and wall temperatures, respectively. When the initial temperature is reduced, the total amount of internal energy in the molten metal is reduced and thus, the solidification time decreases. The amount of heat extracted is proportional to the temperature gradient at the mold wall which increases with a drop in the wall temperature. Thus, the drop in wall temperature reduces the solidification time. Hence, the minimum of solidification time is attained at the bottom left corner (\cref{Fig:Parameter Sweep: Uniform Boundary Temperature Sol Time}). The grain size is governed by the local temperature gradients and cooling rates which are weakly dependent on the initial temperature. Thus, it can be seen that the contour lines are nearly horizontal in \cref{Fig:Parameter Sweep: Uniform Boundary Temperature Max Grain}. On the other hand, as wall temperature reduces, the rate of heat extraction rises and hence, the grains get less time to grow. This causes a drop in the grain size. Thus, the minimum of the maximum grain size is on the bottom right corner (\cref{Fig:Parameter Sweep: Uniform Boundary Temperature Max Grain}). The contour values in \cref{Fig:Parameter Sweep: Uniform Boundary Temperature Min Yield} are negative as maximization objective of the minimum yield strength is converted into minimization by inverting the sign. The minimum is at the top right corner of \cref{Fig:Parameter Sweep: Uniform Boundary Temperature Min Yield}. \Cref{Fig:Parameter Sweep: Split Boundary Temperature} has similar plots for the second case of constant initial temperature and split boundary temperature. \Cref{Fig:Parameter Sweep: Split Boundary Temperature Min Yield} shows the effect of nonuniform boundary temperature. The minimum is attained at wall temperatures of 500 K and 700 K since the local gradients and cooling rates vary due to the asymmetry in the geometry. This analysis shows the utility of the optimization with respect to nonuniform mold wall temperatures.
\Cref{Tab:Single Objective Optimization: Genetic Algorithm Estimates compared with Exact Values} lists the optima for the single objective problems estimated from parameter sweep and GA. Note that since the 200 K range is divided into 200 divisions, the resolution of the parameter sweep estimation is 1 K. For all the six cases, the outputs and corresponding inputs show that the GA estimates are accurate. The GA parameters are varied in the following ranges: [25--100] generations, [10--50] population size, [0.75--0.9] crossover probability and [0.05--0.2] mutation probability. Similar ranges have been used in the literature \cite{kittur2016modeling,gc2016intelligent,gc2017back}. Parameter values of 50 generations with population size of 25 and crossover and mutation probability of 0.8 and 0.1 respectively are found to give accurate estimates as shown in \cref{Tab:Single Objective Optimization: Genetic Algorithm Estimates compared with Exact Values}. The elitist version of GA is used which passes on the best individual from previous generation to the next generation.
\begin{table}[H]
	\centering
	\resizebox{\textwidth}{!}{%
	\begin{tabular}{|c|c|c|c:c|c:c|}
		\hline
		\multirow{2}{*}{\begin{tabular}[c]{@{}c@{}}Problem \\ Type\end{tabular}} & \multirow{2}{*}{\begin{tabular}[c]{@{}c@{}}Optim. \\ Objective\end{tabular}} & \multirow{2}{*}{\begin{tabular}[c]{@{}c@{}}Optim. \\ Type\end{tabular}} & \multicolumn{2}{c|}{Param. Sweep} & \multicolumn{2}{c|}{GA Estimates} \\ \cline{4-7} 
		&  &  & Inputs (K) & Output & Inputs (K) & Output \\ \hline
		\multirow{3}{*}{\begin{tabular}[c]{@{}c@{}}Uniform\\B.C.\end{tabular}} &Solid. Time (s)& Min. &900, 500& 1.962 & 900.2, 500.2 & 1.962 \\ \cdashline{2-7} 
		&Max. Grain ($\mu$m)& Min. & 1100, 500 & 22.41 &1099.1, 500.3& 22.42 \\ \cdashline{2-7} 
		&Min. Yield (MPa)& Max. & 1100, 700 & 145.4 &1099.9, 699.9& 145.4 \\ \hline
		\multirow{3}{*}{\begin{tabular}[c]{@{}c@{}}$T_{init}=$\\1000 K\end{tabular}} &Solid. Time (s)& Min. &500, 500& 1.982 & 500.6, 500.3 & 1.982 \\ \cdashline{2-7} 
		&Max. Grain ($\mu$m)& Min. &500, 500& 22.49 &500.1, 500.1& 22.50 \\ \cdashline{2-7} 
		&Min. Yield (MPa)& Max. & 500, 700 & 140.9 &500.1, 699.9& 140.8 \\ \hline
	\end{tabular}
	}
	\caption{Single Objective Optimization: Genetic Algorithm Estimates compared with Parameter Sweep Values for Two Input Problems}
	\label{Tab:Single Objective Optimization: Genetic Algorithm Estimates compared with Exact Values}
\end{table}
\subsection{Bi-Objective Optimization}
In this section, two objectives are taken at a time for each of the two simplified problems defined in \cref{Sec:Assessment of Genetic Algorithm on Simpler Problems}. As before, a two dimensional mesh of size 40,000 with 200 points in each input dimension is used for this analysis. The outputs are estimated from the neural networks for each of these points. The feasible region is the set of all the attainable designs in the output space which can be estimated by the parameter sweep. For a minimization problem, a design $d_1$ is said to dominate another design $d_2$ if all the objective function values of $d_1$ are less than or equal to $d_2$. The design space can be divided in two disjoint sets $S_p$ and $S_d$ such that $S_p$ contains all the designs which do not dominate each other and at least one design in $S_p$ dominates any design in $S_d$ \cite{deb2001multi}. $S_p$ is called as the Pareto optimal or non-dominated set whereas, $S_d$ is called as the non-Pareto optimal or dominated set. Since the designs in Pareto optimal set are non-dominated with respect to each other, they all are equally good and some additional information regarding the problem is required to make a unique choice out of them. Thus, it is useful to have a list of multiple Pareto optimal solutions. Another way to interpret the Pareto optimal solutions is that any improvement in one objective will worsen at least one other objective thus, resulting in a trade-off \cite{messac2015optimization}. 
\begin{figure}[H]
	\centering
	\begin{subfigure}[t]{0.49\textwidth}
		\includegraphics[width=\textwidth]{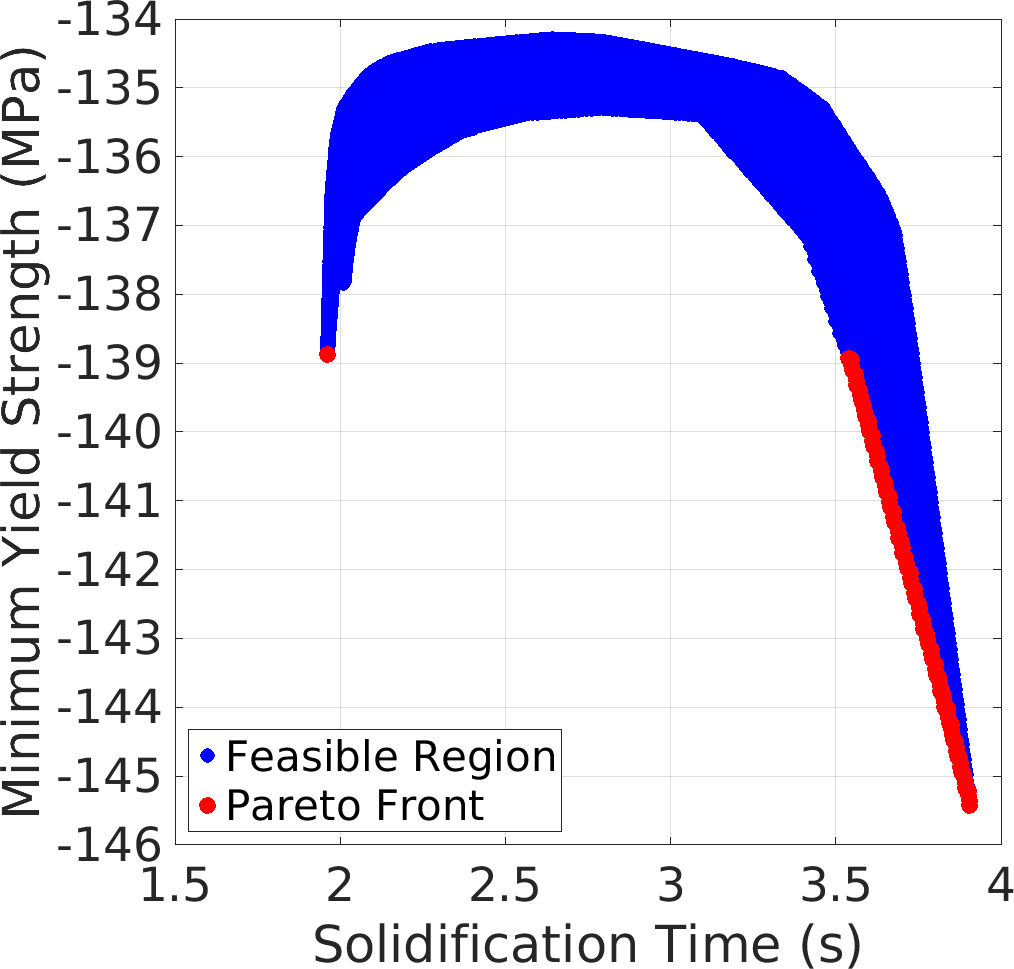}
		\caption{Parameter Sweep}
	\end{subfigure}
	\begin{subfigure}[t]{0.49\textwidth}
		\includegraphics[width=\textwidth]{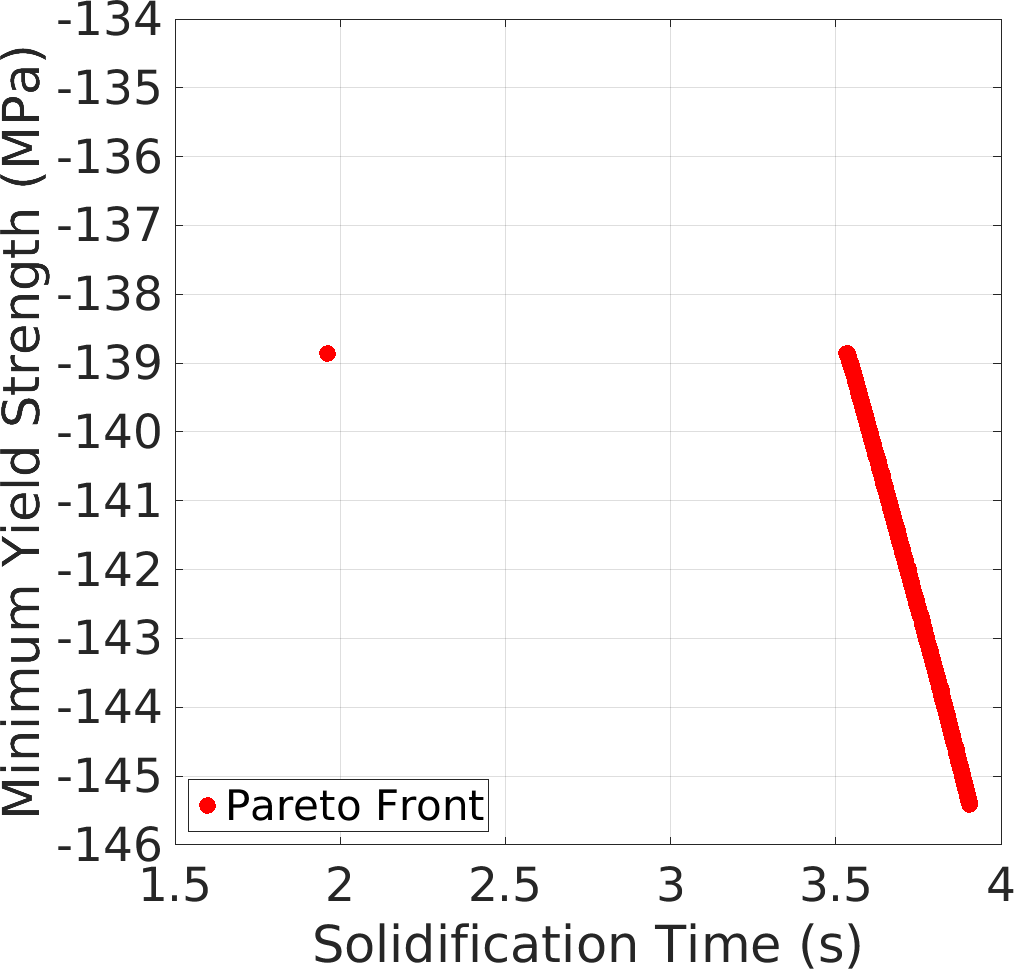}
		\caption{NSGA--II}
	\end{subfigure}	 
	\caption{Uniform Boundary Temperature: Solidification Time vs Min. Yield Strength}
	\label{Fig:Bi-obj Uniform Boundary Temperature: Solidification Time v/s Min. Yield Str.}
\end{figure}
\begin{figure}[H]
	\centering
	\begin{subfigure}[t]{0.49\textwidth}
		\includegraphics[width=\textwidth]{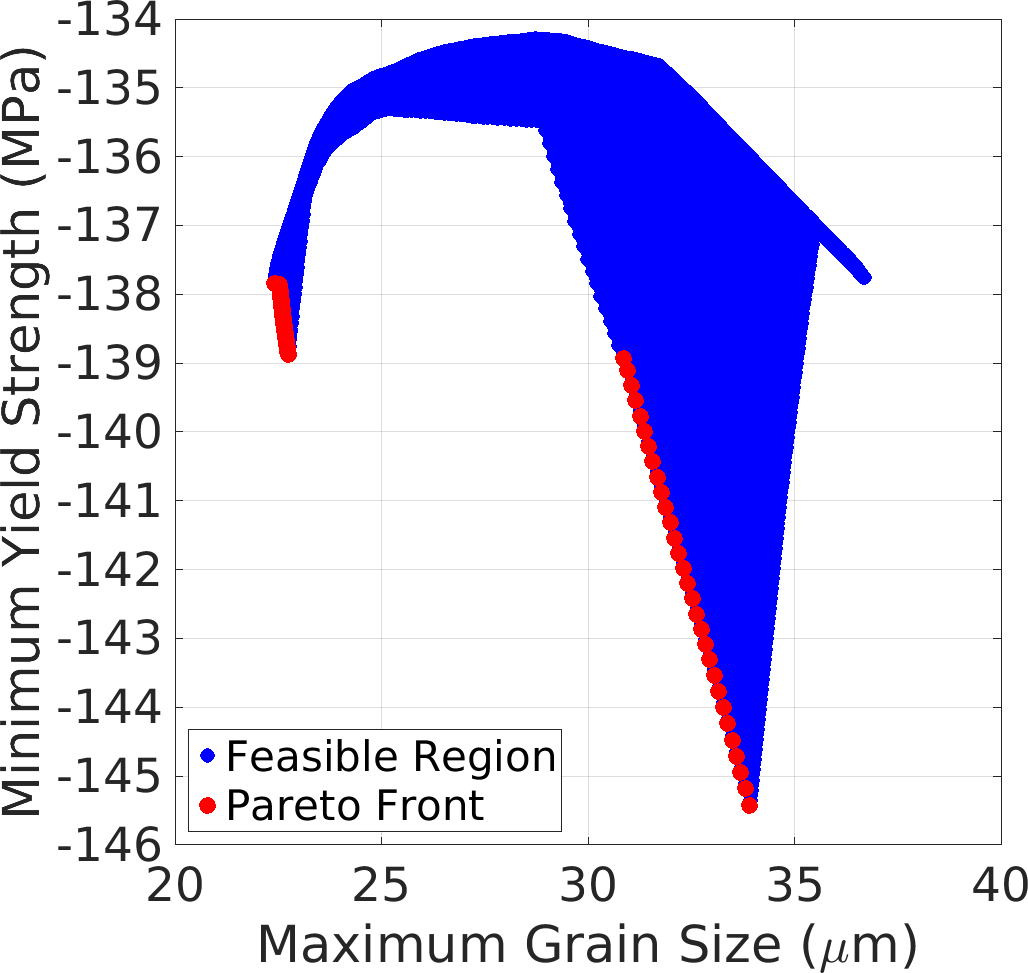}
		\caption{Parameter Sweep}
	\end{subfigure}
	\begin{subfigure}[t]{0.49\textwidth}
		\includegraphics[width=\textwidth]{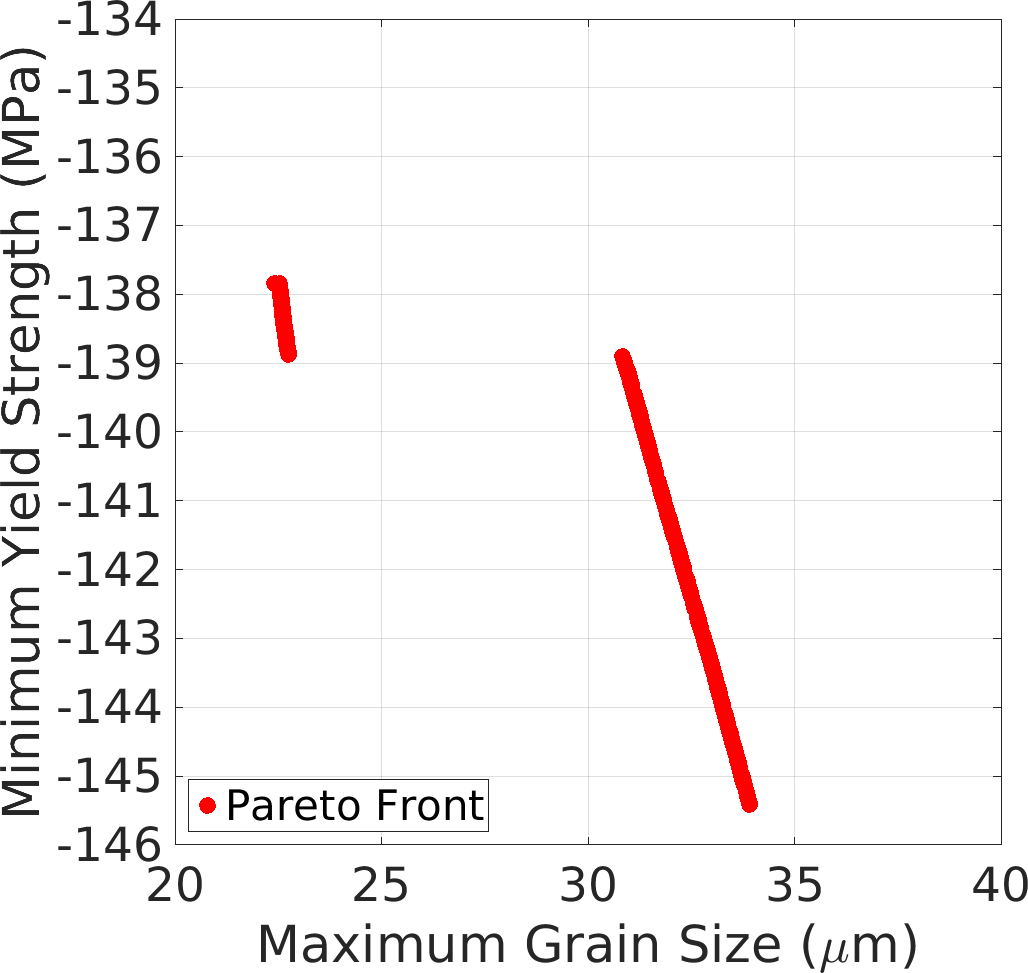}
		\caption{NSGA--II}
	\end{subfigure}	 
	\caption{Uniform Boundary Temperature: Max. Grain Size vs Min. Yield Strength}
	\label{Fig:Bi-obj Uniform Boundary Temperature: Max. Grain Size v/s Min. Yield Str.}
\end{figure}
\par The blue region in the left parts of \cref{Fig:Bi-obj Split Boundary Temperature: Max. Grain Size v/s Min. Yield Str.,Fig:Bi-obj Split Boundary Temperature: Solidification Time v/s Min. Yield Str.,Fig:Bi-obj Uniform Boundary Temperature: Max. Grain Size v/s Min. Yield Str.,Fig:Bi-obj Uniform Boundary Temperature: Solidification Time v/s Min. Yield Str.} indicates the feasible region. Using a pairwise comparison of the designs in the feasible region obtained by parameter sweep, the Pareto front is estimated which is plotted in red. The right side plots of \cref{Fig:Bi-obj Split Boundary Temperature: Max. Grain Size v/s Min. Yield Str.,Fig:Bi-obj Split Boundary Temperature: Solidification Time v/s Min. Yield Str.,Fig:Bi-obj Uniform Boundary Temperature: Max. Grain Size v/s Min. Yield Str.,Fig:Bi-obj Uniform Boundary Temperature: Solidification Time v/s Min. Yield Str.} show the Pareto fronts obtained using NSGA--II. The NSGA parameters are varied in the following ranges: [25--100] generations, [500--1500] population size, [0.75--0.9] crossover probability and [0.05--0.2] mutation probability. The population size used in this work is kept higher than the literature \cite{vundavilli2015neural} to get a good resolution of the Pareto front. It can be seen that both the estimates match which implies that the NSGA--II implementation is accurate. A population size of 1000 is evolved over 50 generations with crossover and mutation probability of 0.8 and 0.1 respectively. Existence of multiple designs in the Pareto set implies that the objectives are conflicting. This can be confirmed from the single objective analysis. For instance, consider the two objectives solidification time and minimum yield strength in the uniform boundary temperature case. From \cref{Fig:Parameter Sweep: Uniform Boundary Temperature Sol Time,Fig:Parameter Sweep: Uniform Boundary Temperature Min Yield} it can be seen that individual minima are attained at different corners. Moreover, the directions of descent are different for each objective and thus, at some points, improvement in one objective can worsen other. This effect is visible on the corresponding Pareto front plot in \cref{Fig:Bi-obj Uniform Boundary Temperature: Solidification Time v/s Min. Yield Str.}.
\begin{figure}[H]
	\centering
	\begin{subfigure}[t]{0.49\textwidth}
		\includegraphics[width=\textwidth]{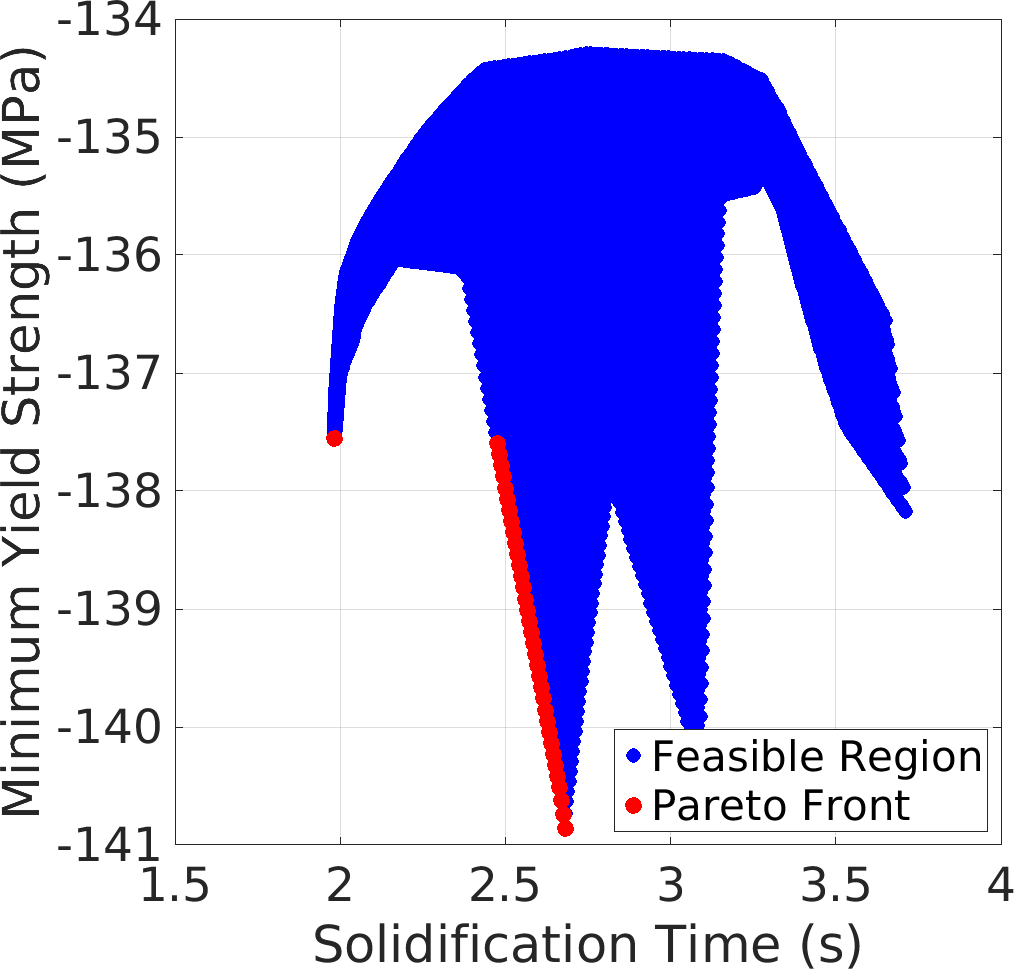}
		\caption{Parameter Sweep}
	\end{subfigure}
	\begin{subfigure}[t]{0.49\textwidth}
		\includegraphics[width=\textwidth]{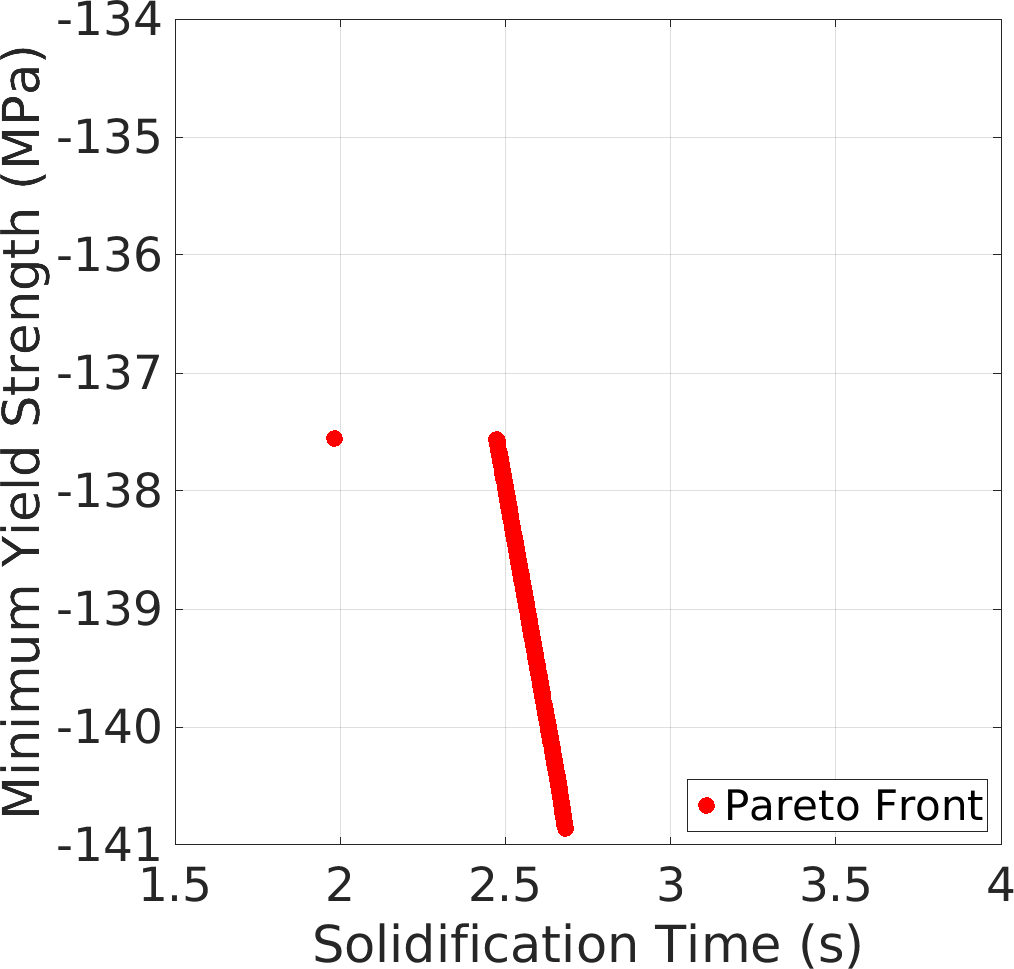}
		\caption{NSGA--II}
	\end{subfigure}	 
	\caption{Split Boundary Temperature with $T_{init}=1000$ K: Solidification Time v/s Min. Yield Str.}
	\label{Fig:Bi-obj Split Boundary Temperature: Solidification Time v/s Min. Yield Str.}
\end{figure}
\begin{figure}[H]
	\centering
	\begin{subfigure}[t]{0.49\textwidth}
		\includegraphics[width=\textwidth]{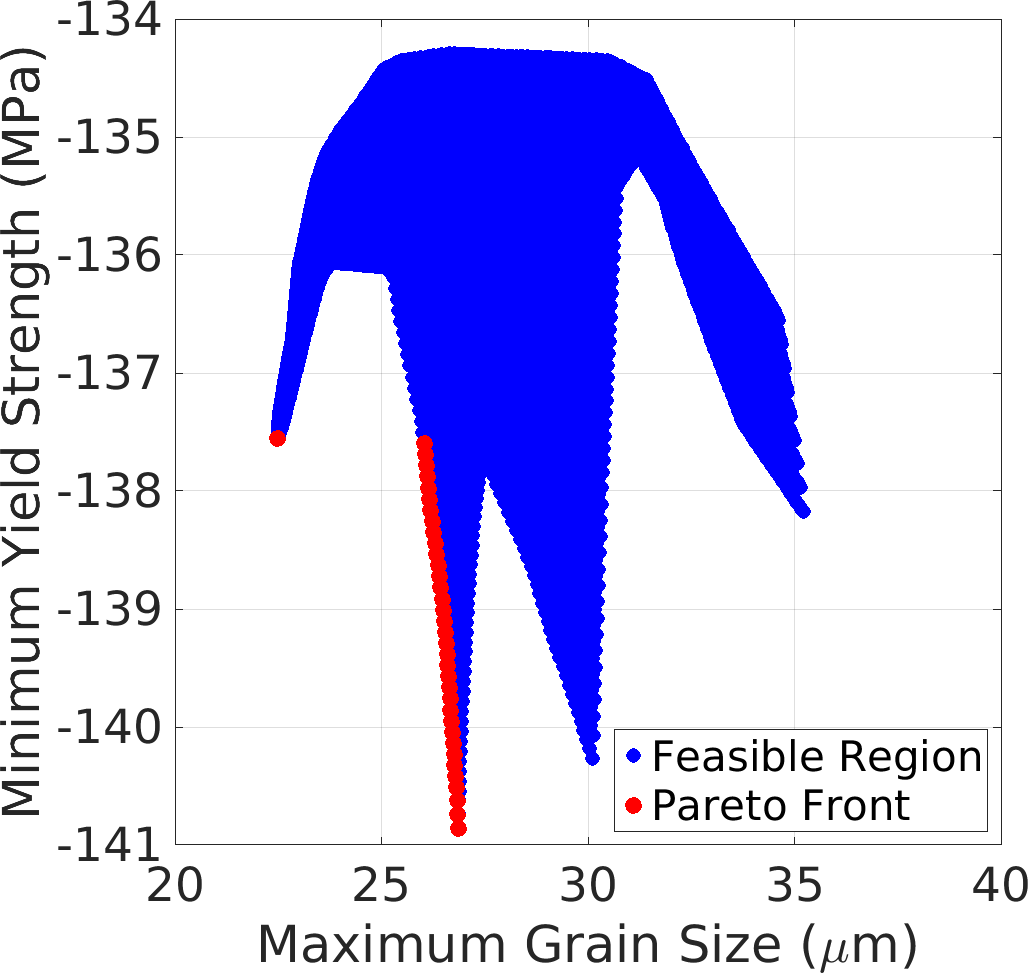}
		\caption{Parameter Sweep}
	\end{subfigure}
	\begin{subfigure}[t]{0.49\textwidth}
		\includegraphics[width=\textwidth]{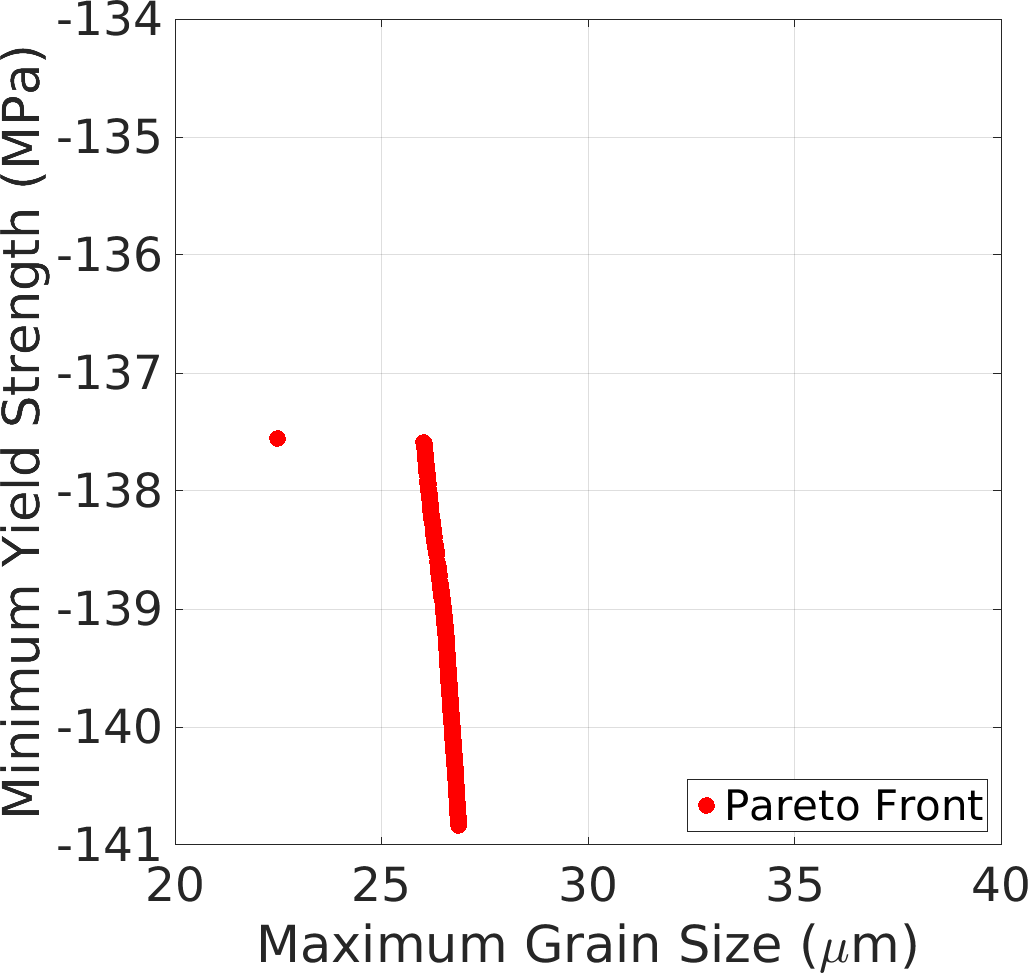}
		\caption{NSGA--II}
	\end{subfigure}	 
	\caption{Split Boundary Temperature with $T_{init}=1000$ K: Max. Grain Size v/s Min. Yield Str.}
	\label{Fig:Bi-obj Split Boundary Temperature: Max. Grain Size v/s Min. Yield Str.}
\end{figure}
\section{Results of Multi--Objective Optimization Problem with Eleven Inputs}
After verification of the NSGA--II implementation on simplified problems, multi--objective design optimization with eleven inputs is solved. As discussed before, some additional problem information is required to choose a single design from all the Pareto optimal designs. In die casting, there is a lot of stochastic variation in the wall and initial temperatures. \citet{shahane2019finite} have performed parameter uncertainty propagation and global sensitivity analysis and found that the die casting outputs are sensitive to the input uncertainty. Thus, from a practical point of view, it is sensible to choose a Pareto optimal design which is least sensitive to the inputs. In this work, such an optimal point is known as a `stable' optimum since any stochastic variation in the inputs has minimal effect on the outputs. A local sensitivity analysis is performed to quantify the sensitivity of outputs towards each input for all the Pareto optimal designs. For a function $\bm{f}: \mathbb{R}^n \rightarrow \mathbb{R}^m$ which takes input $\bm{x} \in \mathbb{R}^n$ and produces output $\bm{f}(\bm{x}) \in \mathbb{R}^m$, the $m \times n$ Jacobian matrix is defined as: 
\begin{equation} 
\mathbb{J}_f[i,j] = \frac{\partial f_i}{\partial x_j} \hspace{1cm} \forall \hspace{0.2cm} 1 \leq i \leq m, 1 \leq j \leq n 
\label{Eq:Jacobian}
\end{equation}
At a given point $\bm{x_0}$, the local sensitivity of $\bm{f}$ with respect to each input can be defined as the Jacobian evaluated at that point: $\mathbb{J}_f (\bm{x_0})$ \cite{smith2013uncertainty}. Here, there are eleven inputs and two outputs. Thus, the $2 \times 11$ Jacobian is estimated at all the Pareto optimal solutions evaluated using the neural networks with a central difference method. Then, the $L_1$ norm of the Jacobian given by the sum of absolute values of all its components is defined as a single scalar metric to quantify the local sensitivity. 
\par
\begin{figure}[H]
	\centering
	\begin{subfigure}[t]{0.49\textwidth}
		\includegraphics[width=\textwidth]{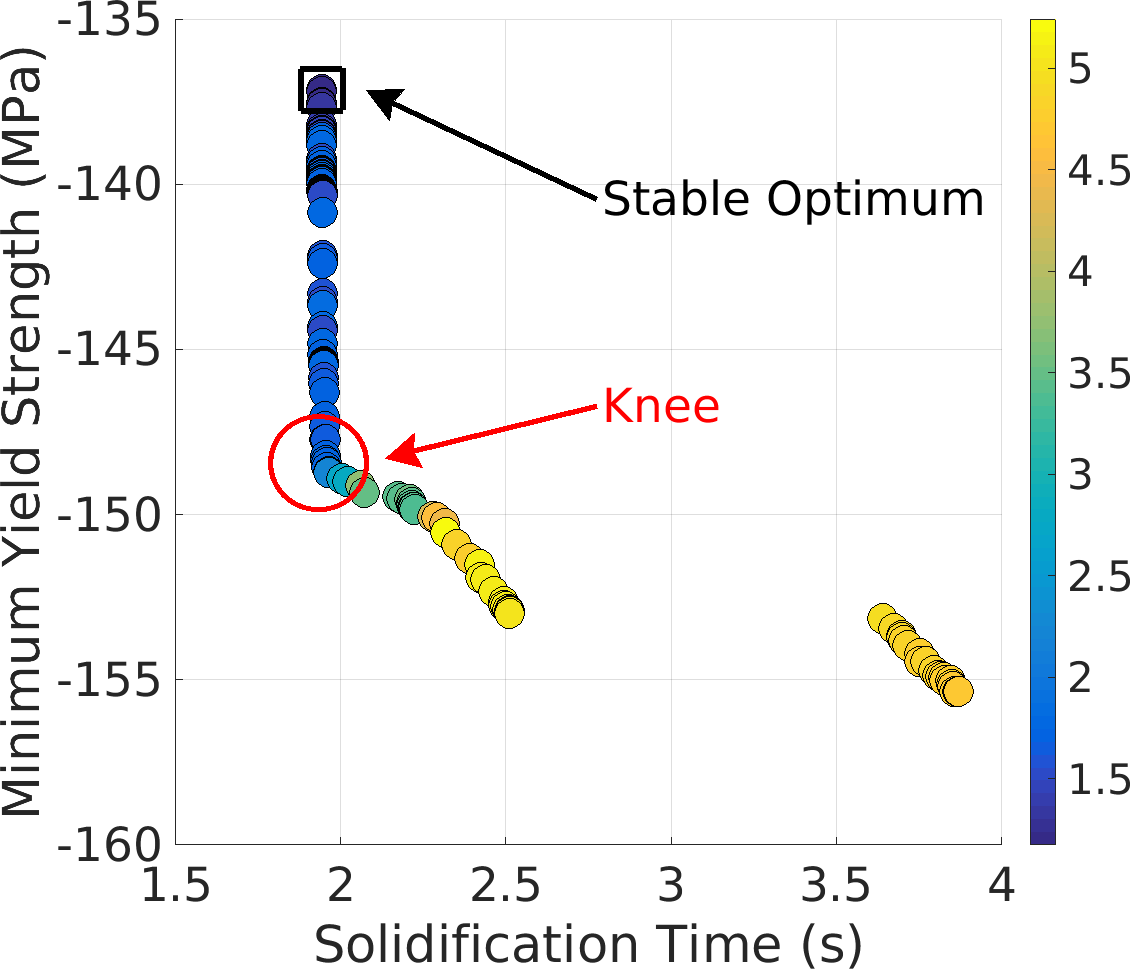}
		\caption{Solidification Time v/s Min. Yield Str.}
	\end{subfigure}
	\begin{subfigure}[t]{0.49\textwidth}
		\includegraphics[width=\textwidth]{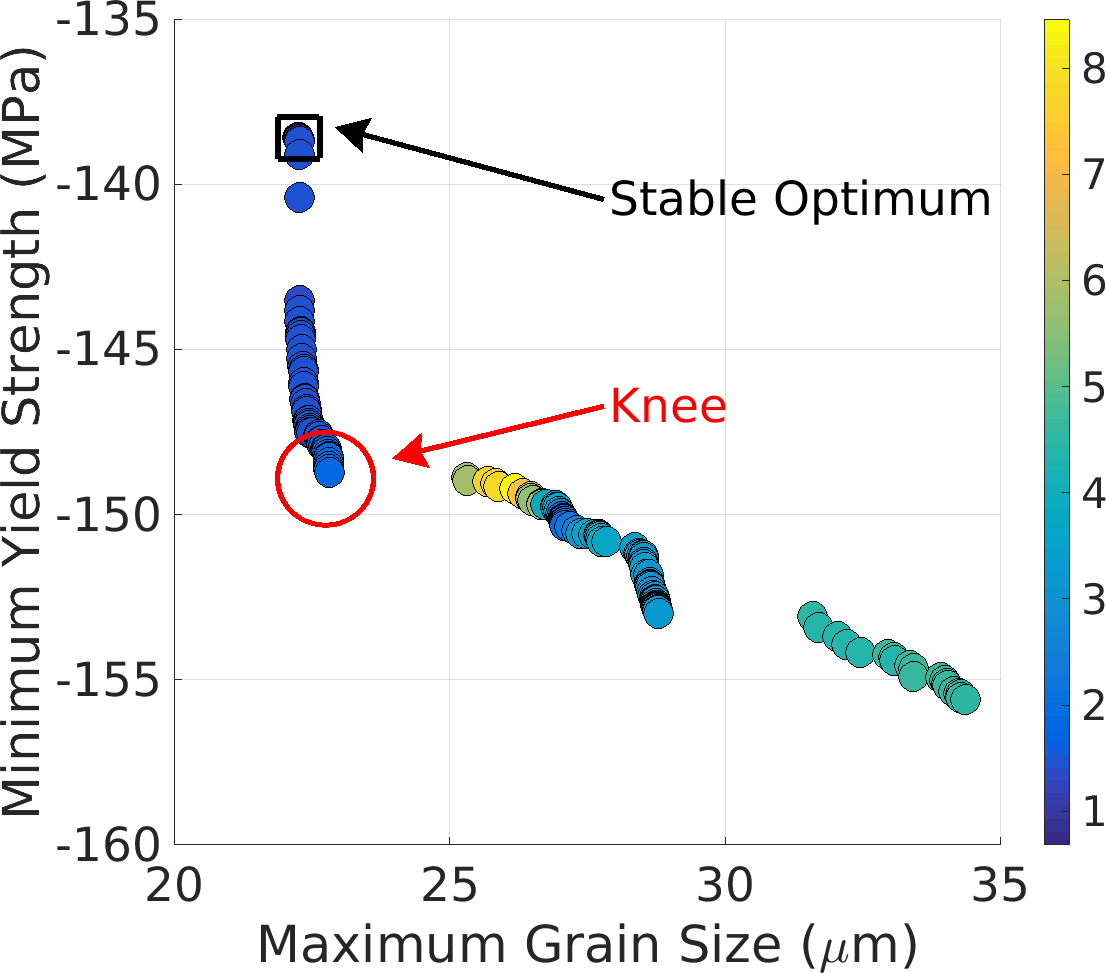}
		\caption{Max. Grain Size v/s Min. Yield Str.}
	\end{subfigure}	 
	\caption{Pareto Front for Two Objectives (Colored by Jacobian Norn)}
	\label{Fig:Pareto Front: 11D}
\end{figure}

 To begin with, two pairs of objectives are chosen: \{solidification time, minimum yield strength\} and \{maximum grain size, minimum yield strength\}. For both of these cases, a population size of 500 with 5000 generations is set. \Cref{Fig:Pareto Front: 11D} plots the Pareto fronts colored by the value of Jacobian norm at each design. It can be seen that the norm varies significantly and thus, ranking the designs based on the sensitivity is useful. The design with minimum norm is chosen and marked on the Pareto fronts as a stable optimum. Note that minimum norm is observed at the end of the Pareto front. However, the norm is low on the entire left vertical side of the Pareto front. Hence, it may be a good idea to choose the design near shown `knee' region which has similar value of the objective on the X--axis but much lower value of the objective on the Y--axis.
\par
\begin{figure}[H]
	\centering
	\includegraphics[width=\textwidth]{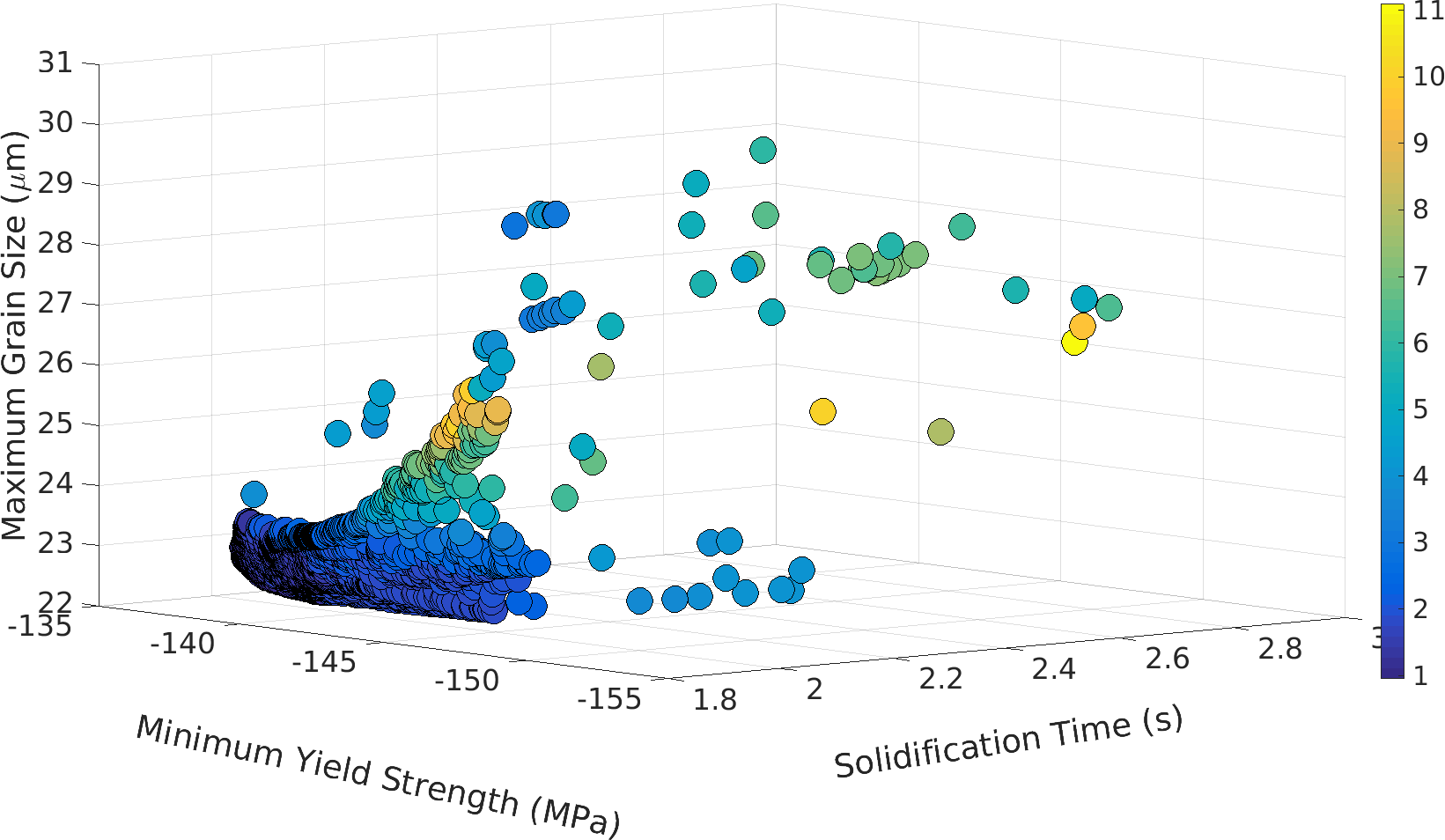}
	\caption{Pareto Front for Three Objectives (Colored by Jacobian Norn)}
	\label{Fig:Pareto Front for 3 Objectives}
\end{figure}

\begin{figure}[H]
	\centering
	\includegraphics[width=0.75\textwidth]{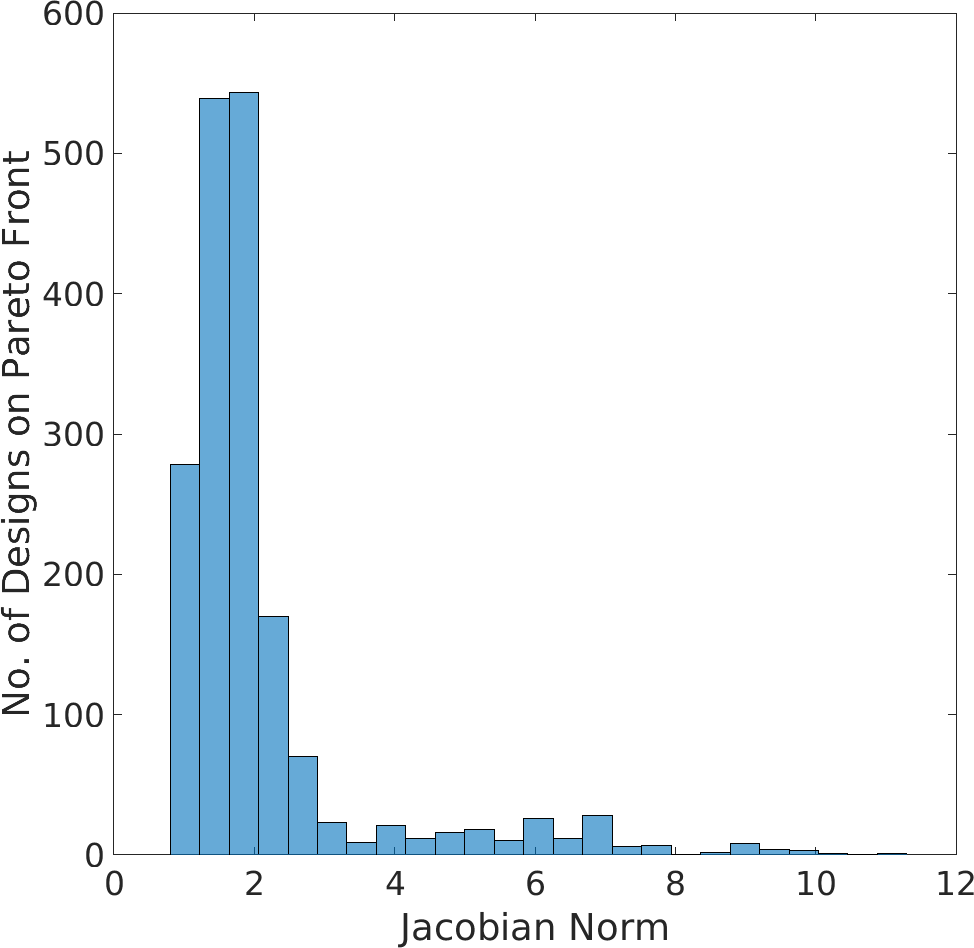}
	\caption{Histogram of Local Sensitivity of Designs on the Pareto Front for 3 Objectives}
	\label{Fig:Local Sensitivity for Designs on Pareto Front for 3 Objectives}
\end{figure}

 The next step is to perform the complete multi--objective optimization analysis as mentioned in \cref{Eq:optim_problem}. NSGA--II is used with a population size of 2000 evolved over 250 generations. \Cref{Fig:Pareto Front for 3 Objectives} plots the Pareto optimal designs in a three dimensional objective space colored by the value of Jacobian norm at each design. Since it is difficult to visualize the colors on the three dimensional Pareto front, a histogram of norm of the Jacobian at each of these designs is also plotted in \cref{Fig:Local Sensitivity for Designs on Pareto Front for 3 Objectives}. It can be seen that the norm varies from 0.95 to 11.1. The histogram is skewed towards the left which implies that multiple designs are stable. The stable optimum is:
\begin{equation} 
	\begin{aligned}
	\text{Inputs: }& T_{init} = 1015.8 \text{ K } \\
	&\bm{T}_{wall} =  \{500.7,502.8,500.0,501.5,500.5,\\&\hspace{2.0cm} 503.6,643.6,508.8,502.3,500.7\}\text{ K}\\
	\text{Outputs: }&  \text{Solidification Time} = 1.99 \text{ s}\\
	&\text{Max Grain Size} = 22.39 \hspace{0.2cm} \mu \text{m}\\
	&\text{Min Yield Strength} = 137.95 \text{ MPa}	
	\label{Eq:Final Optimum Answer}
	\end{aligned}
\end{equation}
\section{Conclusions}
This paper presents an application of multi--objective optimization of the solidification process during die casting. Although the procedure is illustrated for a model clamp geometry, the process is general and can be applied to any practical geometry. Practically, it is not possible to hold the entire mold wall at a uniform temperature. The final product quality in terms of strength and micro-structure and process productivity in terms of solidification time depends directly on the rate and direction of heat extraction during solidification. Heat extraction in turn depends on the placement of coolant lines and coolant flow rates thus, being a crucial part of die design. In this work, the product quality is assessed as a function of initial molten metal and boundary temperatures. Knowledge of boundary temperature distribution in order to optimize the product quality can be useful in die design and process planning. NSGA--II, which is a popular multi--objective genetic algorithm, was used for the optimization process. Since the number of function evaluations required for a GA is extremely high, a deep neural network was used as a surrogate to the full computational fluid dynamics simulation. The training and testing of the neural network was completed with less than thousand full scale finite volume simulations. The run time per simulation using OpenCast was about 20 minutes on a single processor i.e., around 333 compute hours for 1000 simulations. All the simulations were independent and embarrassingly parallel. Thus, a multi--core CPU was used to speed up the process without any additional programming effort for parallelization. Computationally, this was the most expensive part of the process. Subsequent training and testing of the neural network took a few minutes. Implementation of GA is computationally cheap since the evaluation of a neural network is a sequence of matrix products and thus, was completed in few minutes. Hence, it can be seen that the strategy of coupling the GA and neural network with finite volume simulations is computationally beneficial.
\par In this work, the wall is divided into ten domains. Together with the initial temperature, this is an optimization problem with eleven inputs. Both single and multi--objective genetic algorithms were programmed and verified with parameter sweep estimation for simplified versions of the problem. The single objective response surfaces were used to get an insight regarding the conflicting nature of the objectives since the individual optimal solutions were completely different from each other. Moreover, the solidification time, maximum grain size and minimum yield strength varied in the ranges [2, 3.5] seconds, [22, 34] microns and [134, 145] MPa respectively for the given inputs. This showed the utility of the simultaneous optimization of all the objectives since there was a significant scope for improvement. After estimating multiple Pareto optimal solutions, a common question is to choose a single design. The strategy of choosing the design with minimum local sensitivity towards the inputs was found to be practically useful due to the stochastic variations in the input process parameters. Overall, although die casting was used as an example for demonstration, this approach can be used for process optimization of other manufacturing processes like sand casting, additive manufacturing, welding etc. 
\section*{Acknowledgments}
This work was funded by the Digital Manufacturing and Design Innovation Institute with support in part by the U.S. Department of the Army. Any opinions, findings, and conclusions or recommendations expressed in this material are those of the author(s) and do not necessarily reflect the views of the Department of the Army. The authors would like to thank Beau Glim of North American Die Casting Association (NADCA) and Alex Monroe of Mercury Castings for their insightful suggestions.
\section*{References}
\bibliography{References}
\end{document}